\documentclass{aa}

\usepackage{natbib,epsfig,amsmath,amssymb,color}

\def\ga{\,\hbox{\hbox{$ > $}\kern -0.8em \lower 1.0ex\hbox{$\sim$}}\,}
\def\la{\,\hbox{\hbox{$ < $}\kern -0.8em \lower 1.0ex\hbox{$\sim$}}\,}
\def\beq{\begin{equation}}
\def\eeq{\end{equation}}

\titlerunning{Self-similar spiral solutions for accretion disc}
\authorrunning{Hennebelle et al.}

\begin{document}

\title{Spiral-driven accretion in protoplanetary discs - II
Self-similar solutions}

\author{Patrick Hennebelle\inst{1,2}, Geoffroy Lesur \inst{3,4}, S\'ebastien Fromang \inst{1}}
\institute{
Laboratoire AIM, 
Paris-Saclay, CEA/IRFU/SAp - CNRS - Universit\'e Paris Diderot, 91191, 
Gif-sur-Yvette Cedex, France
\and
LERMA (UMR CNRS 8112), Ecole Normale Sup\'erieure, 75231 Paris Cedex, France
\and Univ. Grenoble Alpes, IPAG, 38000, Grenoble, France
\and CNRS, IPAG, F-38000 Grenoble, France}

\abstract
{Accretion discs are ubiquitous in the universe and it is a crucial
  issue to understand how angular momentum and mass are being radially
  transported in these objects.} 
{Here, we study the role played by non-linear spiral patterns within hydrodynamical and non self-gravitating accretion disc 
assuming that external disturbances such as
  infall onto the  disc may trigger them.}
{To do so, we computed self-similar solutions that describe discs
  in which a spiral wave propagates. Such solutions present both
  shocks and critical sonic points that we carefully analyze. }
{For all allowed temperatures and for several
  spiral shocks, we calculated the wave structure. In particular we inferred the
angle of the spiral patern, 
the stress it exerts on the disc as well as the associated flux of
mass and angular momentum as a function of temperature. We quantified 
the rate of angular momentum transport by means of the dimensionless
$\alpha$ parameter. For the thickest disc we considered (corresponding
to $h/r$ values of about 1/3), we found values of $\alpha$ as high as
$0.1$, and scaling with the temperature $T$ such that $\alpha \propto
T^{3/2} \propto (h/r)^3$. The spiral angle scales with the temperature as $\arctan(r/h)$. }
{The existence of these solutions suggests that perturbations occurring 
at disc outer boundaries, such as for example perturbations due to infall motions, can propagate deep inside the disc and
therefore should not be ignored, even when considering small radii. }

\maketitle

\section{Introduction}

Accretion discs play a major role in astrophysics as they are ubiquitously 
observed around stars and black holes. A major question regarding disc
evolution is obviously how angular momentum and mass are transported
radially in reasonably short times \citep[e.g.][]{pringle1981}. It is
widely admitted that local instabilities such as the magneto-rotational
instability \citep[MRI, e.g.][]{balbus2003} or the gravitational
instability \citep[e.g.][]{lodatorice2004} are responsible for
triggering the transport of momentum and mass. Alternatively, 
winds emitted by magnetic processes into the disc may carry away 
angular momentum \citep{turner2014}.  
While it is now widely demonstrated that such instabilities and 
wind launching  lead to
efficient transport, the conditions under which they operate are still
the focus of active research. For example, whether protoplanetary (PP)
discs are always sufficiently ionised is still a matter of debate 
\citep[e.g.][]{lesur+2014}. 
 
The essence of these instabilities is to trigger either non-axisymmetric motions 
or magnetic field configurations within the discs, 
which therefore exert a torque on the gas and lead to an outward flux of 
angular momentum. 
For simplicity reasons, accretion discs have most of the time been
studied in isolation, generally starting with discs at equilibrium. A
notable exception concerns many of the studies that have addressed the
question of disc formation in the context of molecular core
collapse. In that case, the discs are usually massive and
self-gravitating and it is generally admitted that angular
momentum can then be transported by gravitational torques or by
magnetic braking
\citep[e.g.][]{vorobyov2008,machida+2010,joos+2012,li+2013,vorobyov2015}. The
 importance of pressure exerted at the accretion shock  on
the fragmentation of self-gravitating discs within dense cores was
also stressed by \citet{hennebelle+2004} in this context \citep[see also][]{harsono2011}.

The possible specific role that external accretion may have on more
evolved accretion discs has received much less attention and only
a few studies have investigated in details its impact.
This includes in particular the analytical study of \citet{spruit87} who 
computed non-axisymmetric and stationary solutions with shocks  
\citep[see also][]{larson1990, vishniac1989} and the numerical
simulations performed by
\citet{sawada1987,spruit1987b,rozyczka1993,yukawa1997} who studied the 
effect of external accretion on mass and angular momentum transport
within a disc of mass transferring binary system. In this last case,
since accretion is due to the  companion of the accreting object, it
constitutes an obvious cause for the disc to be maintained in a
non-axisymmetric state on long timescales, possibly resulting in
a significant torque within the system. Indeed, these
authors found that accretion is a possible powerful source of symmetry
breaking and results in an efficient transfer of mass.

In the context of PP discs, the impact of accretion has been 
suggested by \citet{padoan+2005,throopbally2008,klessenhennebelle2010,padoan+2014},  
who noted, either by measuring the accretion rate $\dot{M}_{d}$ onto the
disc itself from simulations, or by using analytical arguments, that
$\dot{M}_{d} \propto M_*^2$, where $M_*$ is the mass of the star.  This
relation is very similar to what can be inferred for the accretion
rate $\dot{M}$ onto the young stars themselves
\citep[e.g.][]{muzerolle2005}
 and suggests that accretion onto
PP discs and accretion onto the star are somehow related. 
We note that recent results indicate a weaker relationship
with $\dot{M}_{d} \propto M_*^{1.4}$ \citep{venuti2014}
which therefore may weaken this argument. However, they also find an anti-correlation between 
the accretion rate and the age of the source which is broadly compatible with the idea 
that infall onto the disc may trigger accretion since infall is likely to decrease
with time as well. 
Recently, \citet[][see also \citeauthor{harsono2011,bae2015}
  \citeyear{harsono2011}]{vorobyov2015} performed a series of 2D
simulations of self-gravitating and viscous discs embedded in their
parent cores and showed that the infall of material, and particularly
its specific angular momentum content, has a drastic
impact on their evolution.

In most of these studies, the exact role played by accretion is not
straightforward to identify since other processes (such as
self-gravity and/or explicit viscosity) are generally considered
\citep[e.g.][]{vorobyov2015}. Also, the accretion fluxes that are
considered usually corresponds to rapidly accreting system such as
class-0 or class-I embedded protostars. However, investigating the
role that accretion exactly has by itself, is important because $i)$
it is mandatory to distinguish between the various mechanisms in order
to disentangle their contributions, $ii)$ there are objects for which
the role of other mechanisms remains debated such as for example low mass
protoplanetary discs. It is therefore important to quantify the possible
effect that external accretion could have onto the disc and in
particular whether it could trigger a flux of mass down to the star.

As a first step toward solving that question, we recently performed a
series of 2D numerical simulations \citep[][see also
\citeauthor{bae2015} \citeyear{bae2015}]{Lesur2015}
in order to investigate the transport that is triggered by infalling
material within a disc that is both unmagnetized and 
non self-gravitating. We have found that the infalling flow generates
very significant disturbances at the outer edge of the disc: 
when quantified in terms of the classical $\alpha$ parameter 
(Eq.~\ref{alpha})
 and for typical accretion rate of 
$10^{-7} M_\odot . {\rm yr}^{-1}$, they lead to $\alpha \sim 10^{-2}$. 
Moreover, at small radii, the effective $\alpha$ does not go to zero but 
instead seems to reach a plateau with values of the order of a few
times $10^{-4}$. These important results therefore open up the
possibility that external disturbances  can propagate through the disc
and generate angular momentum transport even at small radii. 

As the physical understanding leading to such a behaviour remains to be
clarified, it is useful to study analytical solutions. This is the aim
of the present paper where we re-visit, and extend, the  self-similar
solutions of non-axisymmetric stationary flows studied by
\citet{spruit87}. In these solutions, there is no explicit viscosity
and the necessary dissipation is provided by shocks. Such solutions provide 
a simple framework and give a strong hint about the physics at play in 
large scale accretion driven discs. 
Section two presents the formalism and the method we used to solve the
equations. Particular emphasis is made of the nature and the role
played by the critical or transonic points present in the flow and
which is playing an important role to understand their mathematical
nature. In section three we study the physical properties of these
solutions as a function of the disc temperature and discuss the
implications. The fourth section concludes the paper. 
  
\section{Self-similar solutions of externally driven accretion}

Following \citet{spruit87}, we look for self-similar solutions that 
could describe the mass and angular momentum flux within a disc that 
would result from a non-axisymmetry induced by an external influence 
such as non-axisymmetric accretion onto the disc. 
The existence of these solutions is important to establish since it
suggests that spiral modes indeed exist and can propagate from large
radii down to small ones. Our solutions, although close
to the case investigated by \citet{spruit87}, are nevertheless different. 
First of all, \citet{spruit87} includes radiation at the surface of the disc 
(assuming an appropriate spatial dependence for the opacity) while we
will restrict the discussion to locally isothermal discs. This is
indeed a more realistic approximations for PP discs and one that has
been used by many authors in numerical simulations. This particular
issue is important since \citet{spruit87} found a significant dependence on
$\gamma$, the adiabatic index, and it is thus important to clarify the
effect that the effective equation of state has on the  
solutions behaviour. 
In particular in the limit where $\gamma \rightarrow 1$, 
 the spiral angle seems to converge toward a value close to 90 degrees (see Fig.~2 of \citet{spruit87}). 
Since PP discs are  typically locally isothermal one wonders whether such a mode 
could develop and whether it could lead to significant transport.
Second, we explicitly give the dependence of various quantities as a function 
of the gas temperature, while \citet{spruit87} focused onto the dependence on 
$\gamma$. Third, we find that there are two \citep[instead of one as
found by][]{spruit87} possible choices 
for the disc surface density radial profiles. While similar in nature, the two family of solutions 
differ in an important manner, since they respectively correspond to
the two limiting cases of vanishing angular momentum flux and vanishing
mass flux. Fourth, we clarify the mathematical nature of 
the solutions and in particular the topology of the critical points, which 
play a key role for these solutions and is important to solve the 
equations numerically. Finally, we describe a simple
method to obtain these solutions which may serve as reference 
to compare to simulations and observational results. 

\subsection{Ordinary equations for self-similar solutions}

The equations we solve are the usual fluid equations. Written in 
Cylindrical geometry, averaged along the z-direction over the disc scale
height, $h$, and assuming stationarity, they write:

\begin{eqnarray}
\label{eqfluid1}
 \left( u_r \partial_r u_r + {u_\phi \over r} \partial_\phi  u_r  - {u_\phi^2 \over r} \right) = - { \partial _r P \over \Sigma} +  g_r, \\
\label{eqfluid2}
 \left( u_r \partial_r u_\phi + {u_\phi \over r} \partial_\phi  u_\phi  + {u_r u_\phi \over r} \right) = - {\partial _\phi P \over r \Sigma } , \\
\label{eqfluid3}
 {1 \over r } \partial _r ( r \Sigma u_r) + {1 \over r} \partial_\phi ( \Sigma  u_\phi )  = 0, \\
\label{eqfluid4}
g_r = -{G M \over r^2} = -  \Omega^2 r, \\ 
\label{eqfluid5}
h \simeq  { C_s \over \Omega }.
\end{eqnarray}
The disc is assumed to be locally isothermal meaning that $P = C_s(r)
^2 \Sigma$, that is to say  
the sound speed, $C_s$, and the temperature, $T$, depend only on the
radius, $r$. Note that  
$P$ is the vertically averaged pressure and $\Sigma \simeq 2 h \rho$
is the column density
and $\rho$ is the midplane density.
All the other quantities have their usual meaning.
Since no explicit dissipation is considered here, the solutions must necessarily entail
shocks, which will then lead to finite energy dissipation. Indeed, angular momentum 
and mass transport, imply such energy dissipation.

To normalise the system, we use similar conventions as \citet{spruit87}, namely 
\begin{eqnarray}
\nonumber
r = r_0 x, \\
\nonumber
u_r = r_0 \Omega_0 \tilde{u}_r, \\
\label{normalisation}
u_\phi = r_0 \Omega_0( x^{-1/2} +  \tilde{u}_1), \\
\nonumber
\Sigma = \Sigma_0 \tilde{\Sigma}, \\
\nonumber
P = \Sigma_0 r_0^2 \Omega_0 ^2 \tilde{T}   \tilde{\Sigma}, 
\end{eqnarray}
where $r_0$ and $\Sigma_0$ are arbitrary radius and surface density,
$\Omega_0= (G M / r_0^3)^{1/2}$.  
In particular, $r_0 \Omega_0 x^{-1/2}$ is simply the Keplerian velocity.

To obtain self-similar solutions, we introduce a new angular  variable, $\psi = \phi + \beta(x)$, 
that is to say the new angular variable shifts with respect to $\phi$ when $r$ varies
and we look for solutions that can be written as $f(x',\psi)$ instead of $f(x,\phi)$.
Using the definitions stated by Eqs.(\ref{normalisation}), Eqs.~(\ref{eqfluid1}-\ref{eqfluid3}) become
\begin{eqnarray}
\label{eqfluid_b1}
  \tilde{u}_r \partial_x \tilde{u}_r +  \tilde{u}_r \beta' \partial_\psi \tilde{u}_r +  \left( x^{-3/2} + {\tilde{u}_1 \over x} \right) \partial_\psi  \tilde{u}_r  - 
{ \tilde{u}_1^2 \over x}  - && \nonumber \\
  x^{-2} - 2 x^{-3/2} \tilde{u}_1    = - {1 \over \tilde{\Sigma}} \left( \partial _x (\tilde{T} \tilde{\Sigma} \right)  +  \beta ' \partial _\psi (\tilde{T} \tilde{\Sigma}) )
- {1 \over x^2}  &&, \\
\label{eqfluid_b2}
 \tilde{u}_r \partial_x \tilde{u}_1 + \beta ' \tilde{u}_r \partial_\psi \tilde{u}_1 + \left( {1 \over x^{3/2}} + {\tilde{u}_1 \over x} \right) 
\partial_ \psi  \tilde{u}_1  + {\tilde{u}_r \tilde{u}_1 \over x} + && \nonumber \\ 
 {\tilde{u}_r  \over 2 x^{3/2}}  = 
- {1 \over x \tilde{\Sigma} }  \partial _\psi (\tilde{T} \tilde{\Sigma}) , && \\
\label{eqfluid_b3}
 {1 \over x } \partial _x ( x  \tilde{\Sigma} \tilde{u} _r) + \beta ' \partial _\psi (  \tilde{\Sigma} \tilde{u} _r) + {1 \over x} 
\partial_\psi ( \tilde{\Sigma}  \tilde{u}_\phi )  = 0 \, .
\end{eqnarray}

Finally, we seek for self-similar solutions in the radius, $x$ and we set
\begin{eqnarray}
\nonumber
\tilde{u}_r &=& x^{-1/2} U(\psi), \\
\nonumber
\tilde{u}_1 &=& x^{-1/2} V(\psi), \\
\label{selfsim}
\tilde{\Sigma} &=& x^{-n} R(\psi), \\
\nonumber
\tilde{T} &=& x^{-1} T_0, \\
\nonumber
\beta' &=& B x^{-1}. 
\end{eqnarray}
The parameter $B$, which is equal to $r \partial_r \psi $ represents the tangent of $\theta$, 
the angle between the spiral pattern and the radial direction.  
Plunging these expressions into Eqs.~(\ref{eqfluid_b1}-\ref{eqfluid_b3}), we get 
\begin{eqnarray}
\label{eqfluid_od1}
     (B U + 1+V) U' +  B T_0 { R' \over R }    && \\
\nonumber
    =  (n+1) T_0   + {1 \over 2} U^2 + V^2 + 2 V, && \\
\label{eqfluid_od2}
       (B U + 1 + V) V' +  T_0  {R' \over R}   = - {1 \over 2}U (V+1) , \\
\label{eqfluid_od3}
  (-n + {1 \over 2}) R U  +   ( R( B  U + 1 + V ))' = 0.
\end{eqnarray}

It is convenient to introduce the variables 
\begin{eqnarray}
\label{def_W}
W = {B U + 1 + V \over \sqrt{1 + B^2} }, 
\end{eqnarray}
which represents the velocity component normal to the 
spiral pattern and to the shock wave and 
\begin{eqnarray}
\label{def_Z}
Z = {U - B( 1 + V) \over \sqrt{1 + B^2} }, 
\end{eqnarray}
which represents the velocity component parallel to the 
 shock wave.

With these definitions, Eq.~(\ref{eqfluid_od3}) can be rewritten as 
\begin{eqnarray}
\label{eqfluid_od3b}
{R ' \over R} = - {W' \over W} +  \left( n - {1 \over 2} \right) {U \over \sqrt{1+B^2} W},
\end{eqnarray}
and easily combines with Eqs.~(\ref{eqfluid_od1}-\ref{eqfluid_od2}) 
leading to
\begin{eqnarray}
\label{ef_new_od1}
      W'  = {W \over 2 (1+B^2) (W^2 -T_0) }   &&   \times  \nonumber \\
    \left( B W^2 + 2 B Z ^2 \right. - WZ &+& \left.  B \left( 2 (n+1)
    T_0-2 \right) \right. \nonumber\\
&-& \left. T_0 \left(2 n - 1 \right)  {B W + Z \over W} \right),  \\
\label{ef_new_od2}
     Z'  =  {1 \over  (1+B^2) W} && \times \nonumber \\
\big( W^2 + {Z^2 \over 2} - {B \over 2} W Z   &+&  \left( (n+1)  T_0
-1 \right) \big) \, .
\end{eqnarray}
Equations~(\ref{ef_new_od1}-\ref{ef_new_od2}) are ordinary equations of $W$ and $Z$. They present 
a critical point at $W =  \sqrt{T_0}$, that is to say when the velocity perpendicular
to the spiral pattern is equal to the sound speed. As discussed in the next session, 
the critical point, which must be crossed smoothly,  plays an important role to obtain these
solutions. 

\subsection{Boundary conditions}
The boundary conditions are $2\pi$ periodic. However, we allow the disc to have several identical spiral arms, so we ask our solutions to be $2\pi/m$-periodic, where $m$ is the number of spiral arms of the solution.

\subsection{Conserved quantities}
Conservation of mass and momentum in the disc leads to several
constrains that have to be satisfied by the solutions. These
constrains are of two types: jump conditions and integral
conditions. These two types of conditions express
the continuity of mass and momentum fluxes.

\subsubsection{Jump conditions}
As discussed above, since the solutions ought to describe transport of angular momentum
and mass, there is unavoidably energy dissipation in the process. Because the equations do
not entail any viscous terms, it implies that shocks must be present. Therefore
the present solutions must satisfy Rankine-Hugoniot conditions through the shock
that is to say the flux of mass and momentum must be continuous. 
This leads to 
\begin{eqnarray}
\label{RK1}
R_1 W _1 = R_2 W_2, \\
\label{RK2}
Z_1 = Z_2, \\
\label{RK3}
R_1 W_1^2 + R_1 T_0 = R_2 W_2^2 + R_2 T_0.
\end{eqnarray}
The last expression can be replaced 
by the relation 
\begin{eqnarray}
\label{RK_W}
W_1 W_2 = T_0,
\end{eqnarray}
where the subscripts 1 and 2 represent the pre- and post-shock material.  
To obtain this last relation, we can simply 
combine Eq.~(\ref{RK1}) and Eq.~(\ref{RK3}).

\subsubsection{Integral conditions}

Our disc model exhibits two important conservation equations, the conservation of mass (Eq.~\ref{eqfluid_od3})
and the conservation of angular momentum
\begin{eqnarray}
 {1 \over r } \partial _r \left( r (\Sigma r u _\phi) u_r \right) + {1 \over r} \partial_\phi \left(  (\Sigma  r u_\phi ) u_\phi \right) 
= - \partial_\phi P,
\end{eqnarray}
which can easily be obtained by combining Eqs.~(\ref{eqfluid2}-\ref{eqfluid3}) and where,
 as before, stationarity is assumed and integration through the disc is performed.
Using the self-similar variables, we get
\begin{eqnarray}
\label{eqfluid_od4}
   (-n + 1 ) R U (1+V)  &+&   \\ 
( R (1+V) &\times& ( B  U + 1 + V ) + T_0 R)' = 0.
\nonumber
\end{eqnarray}

Integrating Eqns.~(\ref{eqfluid_od3}) and (\ref{eqfluid_od4}) between
0 and $2 \pi/m$, and making use of the periodic boundary conditions, 
we obtain the constraints 
\begin{eqnarray}
\label{eq:cst_mass}	 (-n + 1/2) \int_0 ^{2 \pi/m} R U d\psi &=&0\\
\label{eq:cst_mom} (-n + 1 ) \int _0 ^{2 \pi/m}R U (1+V) d\psi &=&0
\end{eqnarray}

This implies that either the azimuthally averaged mass flux $\int_0
^{2 \pi/m} R U d\psi$ or the angular  
momentum flux $\int _0 ^{2 \pi/m}R U (1+V)$ has to be zero. Although solutions with zero mass and angular 
momentum flux are \emph{a priori} possible for arbitrary $n$ (but likely do not satisfy Eq.~\ref{RK1}), we will focus on more physical solutions: 
constant mass flux solutions with $n=1/2$ (which corresponds to the choice made in Spruit 1987) or 
constant angular momentum flux solutions with $n=1$.  Note that
since the temperature is proportional to $1/r$, this implies  
that the disc thickness, $h \simeq C_s / \Omega \propto r$. Thus 
the density profile is $\rho \propto r^{-(n+1)}=r^{-3/2}$ for $n=1/2$
and $\rho \propto r^{-2}$ for $n=1$.

In the constant mass flux case ($n=1/2$), we see from
Eq.~(\ref{eqfluid_od3}) that 
\begin{eqnarray}
\label{ef_new_od3}
   R W  = K,
\end{eqnarray}
where $K$ is an arbitrary constant. In other words, 
the density variable $R$ is inversely proportional to the velocity
component perpendicular to the shock. This automatically ensures that
Eq.~(\ref{RK1}) is satisfied. 
Note that since the present solutions are stationary, a mass flux through the disc
implies that the central mass increases with time (as will be seen the flux is inwards, 
as expected). 

On the other hand, solutions with $n=1$ present a flux of angular momentum through the disc but 
no flux of mass. This implies that while the particle fluids move on closed orbits, angular momentum 
is transported radially during one cycle. Therefore this implies that a source of angular 
momentum must be present in the centre. What this source exactly represents physically
can be debated. One possibility is that these solutions could represent regimes, limited in time, 
during which the inner part of the disc is providing the corresponding
amount of angular momentum. 
This is typically what happens in the context of the so-called dead disk as emphasized by 
\citet{syunyaev1977} and \citet{dangelo2012}, which can arise when a central object is magnetically coupled to the disc
and exchange angular momentum with it.
While generally the solutions are time-dependent, in some circumstance a stationary regime has been inferred.
The solutions however make the assumption of an axisymetric disc and use the $\alpha$ modelling. The present 
solutions may therefore offer a complementary description in which the mechanisms responsible for the angular momentum
transport through the disc is explicitly described. It is also interesting to note that 
\citet{syunyaev1977} predict the column density of the dead disc to have $n=11/10$ while it has $n=1$ in our case.
Similarly, the steady $\alpha$ disc has $n=3/4$ while it is $n=1/2$ for the self-similar non-axisymetric disc.
It is likely the case that for different values of $n$, the disc is not stationary \footnote{An interesting unsolved question 
is whether different values of  $n$ would correspond to unstationary discs whose average momentum and mass fluxes
could be reasonably described by the expression we obtain here. Indeed as shown later the effective $\alpha$ we get
does not change much between $n=1/2$ and $n=1$ and it is therefore tempting to assume that it could also be the case for $n$ 
not too different from these 2 values.}. These solutions would be limited in time since the amount of angular momentum 
that a star possesses is fairly limited. In this respect, another possible interesting application could be the circumbinary discs 
\citep{dubus2002}.

Using Eq.~(\ref{eqfluid_od4}) and $n=1$,
we get 
\begin{eqnarray}
\label{ef_new_od4}
   R   = {K \over W (W - B Z) + T_0}.
\end{eqnarray}
With this expression, it is easy to show that Eq.~(\ref{RK1}) is 
automatically satisfied when conditions (\ref{RK2}) and (\ref{RK3})  
are valid. Therefore  these solutions are physically meaningful, at
least in this respect. \\

\subsection{The critical point}

The other constraint comes from the critical point that must be crossed
smoothly. Since the existence of shocks connecting supersonic and subsonic regions
 is necessary, and since the solutions are $2\pi/m$ periodic, there must be 
a smooth transition between the subsonic and the supersonic regions implying 
that crossing the critical point is unavoidable.
Mathematically, this implies that both the numerator and the
denominator of Eq.~(\ref{ef_new_od1}) must vanish at that location,
leading to the two conditions  
\begin{eqnarray}
\label{cond_crit1}
W_c = \pm \sqrt{ T_0} , \\
\label{cond_crit2}
B W_c^2 + 2 B Z_c ^2  - W_c Z_c + B(3 T_0-2) = 0, 
\end{eqnarray}
which gives:
\begin{eqnarray}
\label{Z_crit2}
Z_c = {1 \over 2 B} \left( n \sqrt{T_0} \pm 2 \sqrt{{n ^2 T_0 \over 4}
  + B^2 - 2 B^2 T_0} \right) \, .
\end{eqnarray}
This last equation reveals in particular that the condition 
\begin{eqnarray}
T_0 < {1 \over 2} \; {\rm or} \; B^2 > {-n^2 T_0 \over 4 (1- 2 T_0 )}.
\end{eqnarray}
must be satisfied for $Z_c$ to be real. Since discs are flat object,
$h/r \simeq \sqrt{T_0}$ is expected to be small. Therefore, in 
practice, this condition does not restrict the values 
of $B$ and $T_0$.

The  consequence of the existence of the critical point is that in most of the possible range of
parameters, the critical point constitutes a constraint that must be
fulfilled, thereby reducing by one the number of degrees of freedom
of the system. Note that strictly speaking a careful study of its 
topology is actually required to understand exactly the constraints 
it  brings to the system  (see appendix \ref{app:topo}).

\setlength{\unitlength}{1cm}
\begin{figure*} 
\begin{picture} (0,20)
\put(0,16){\includegraphics[width=8cm]{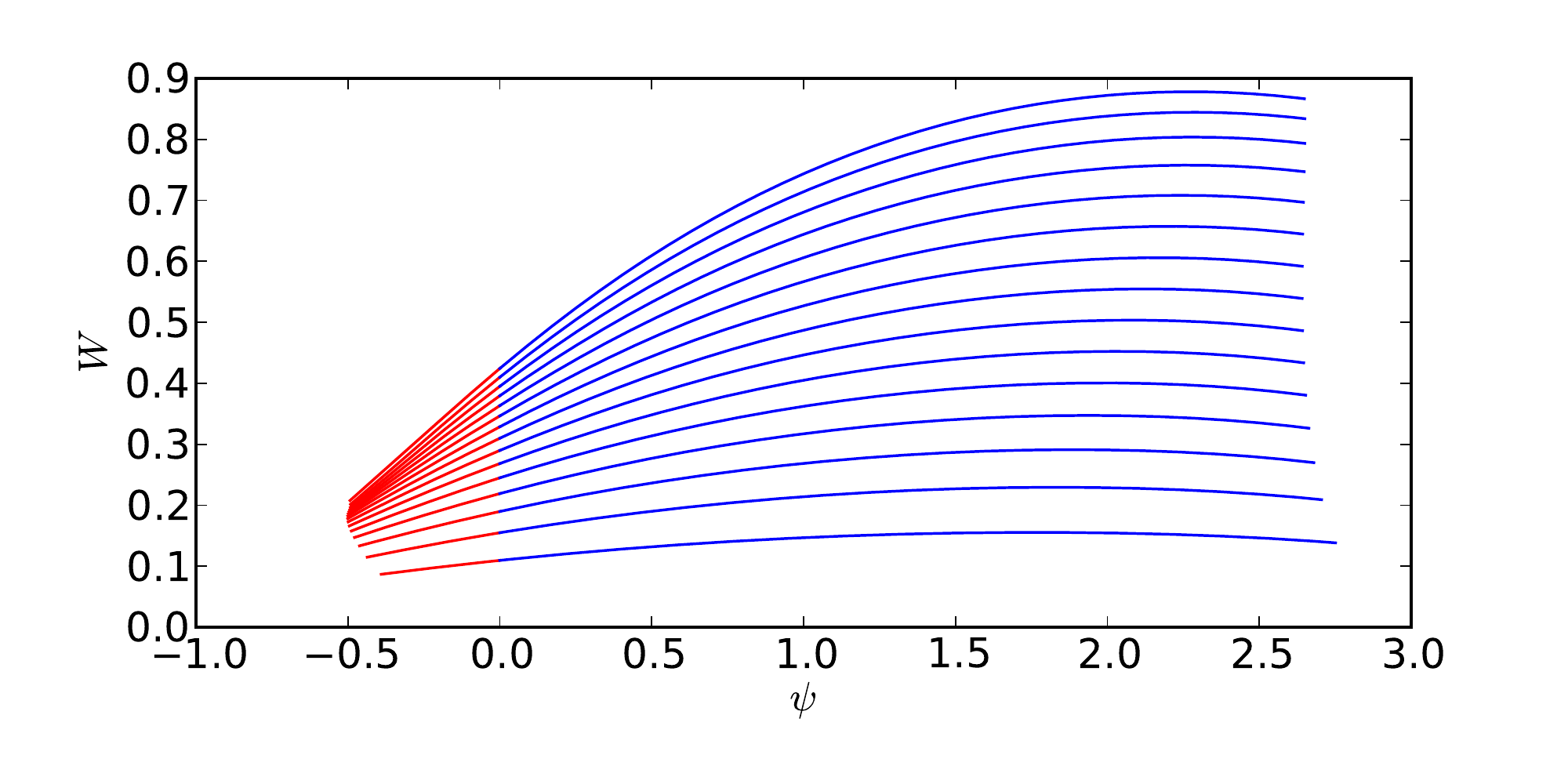}}  
\put(0,12){\includegraphics[width=8cm]{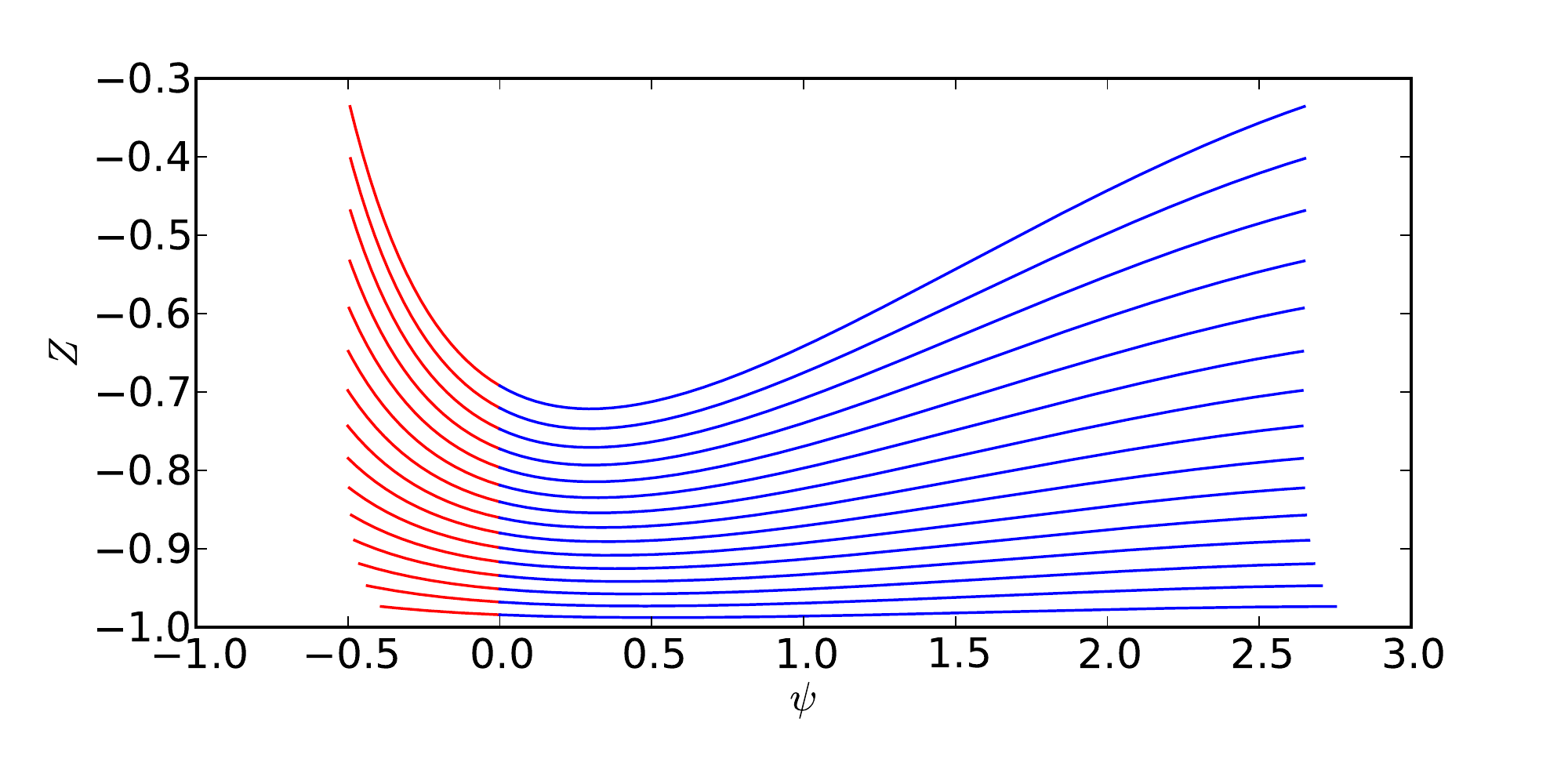}}  
\put(0,8){\includegraphics[width=8cm]{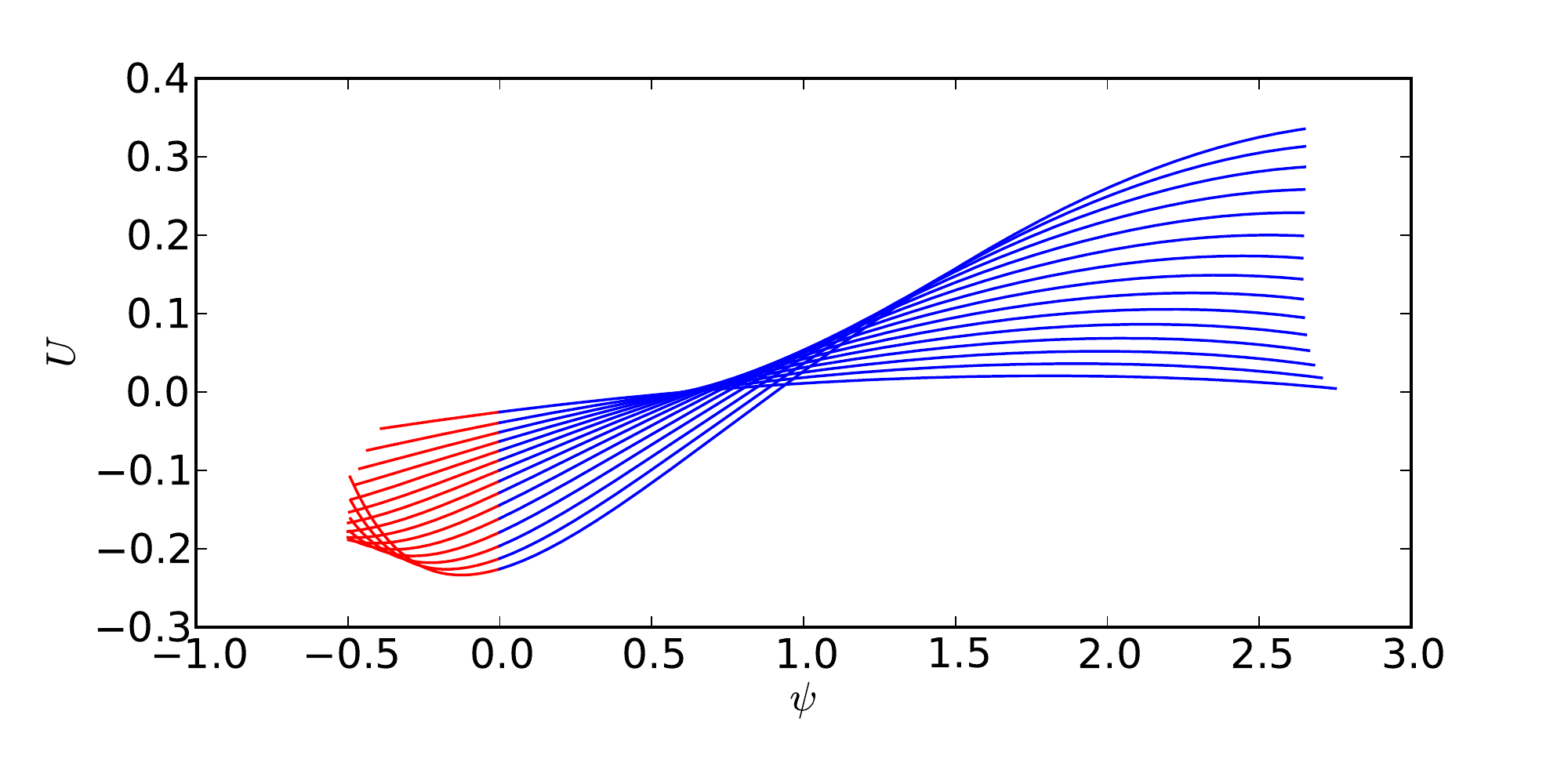}}  
\put(0,4){\includegraphics[width=8cm]{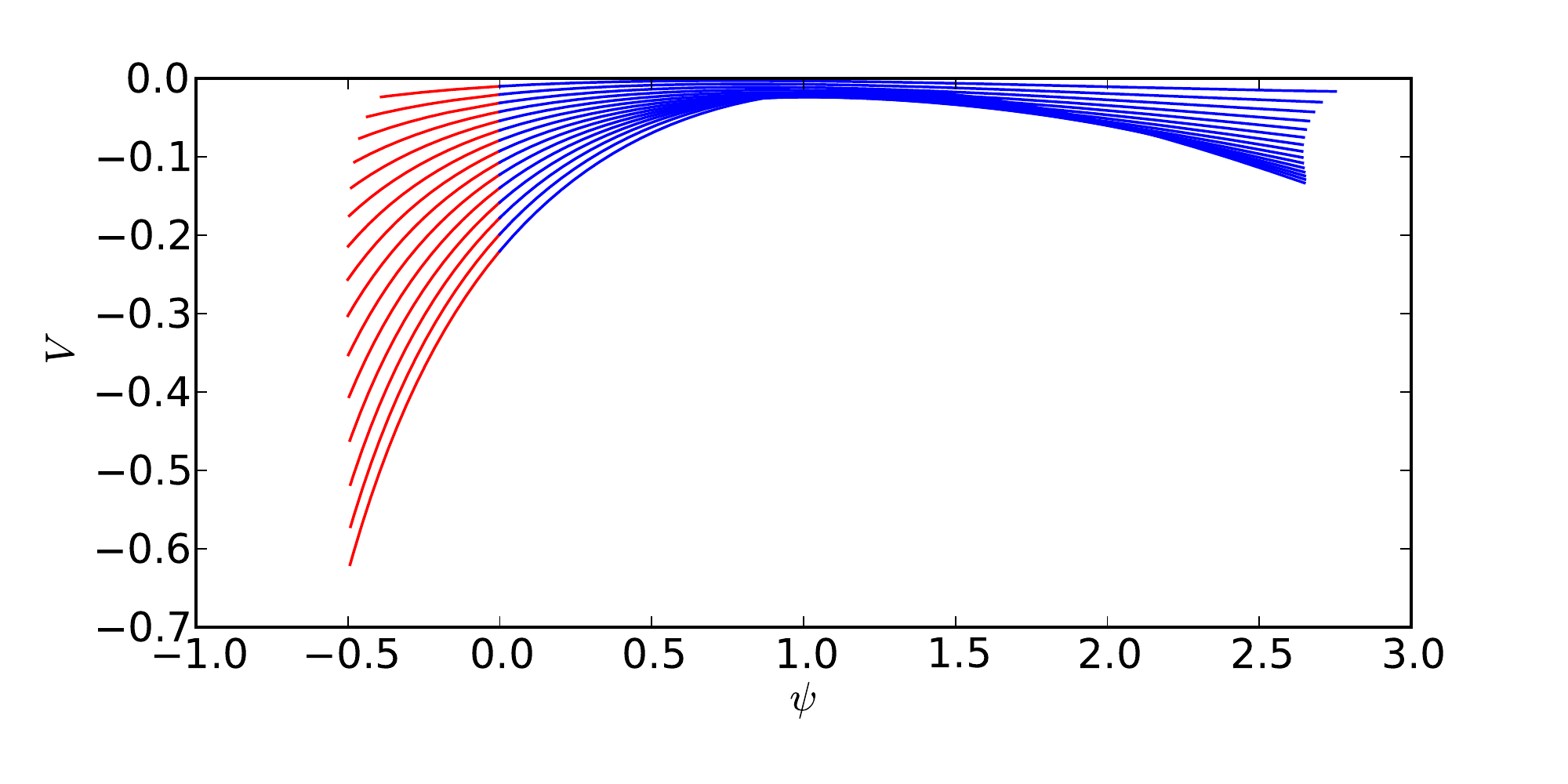}}  
\put(0,0){\includegraphics[width=8cm]{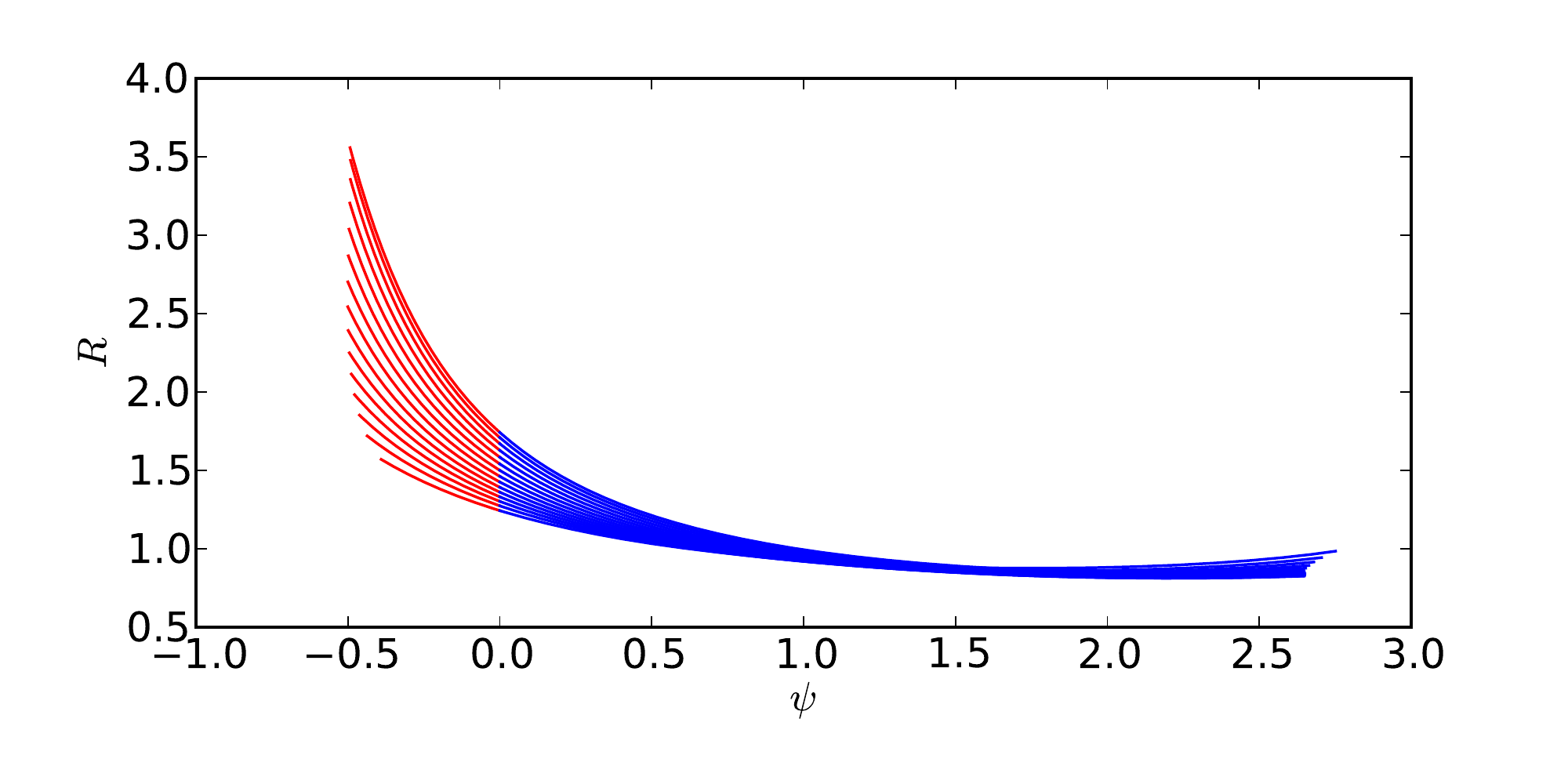}}  
\put(8,16){\includegraphics[width=8cm]{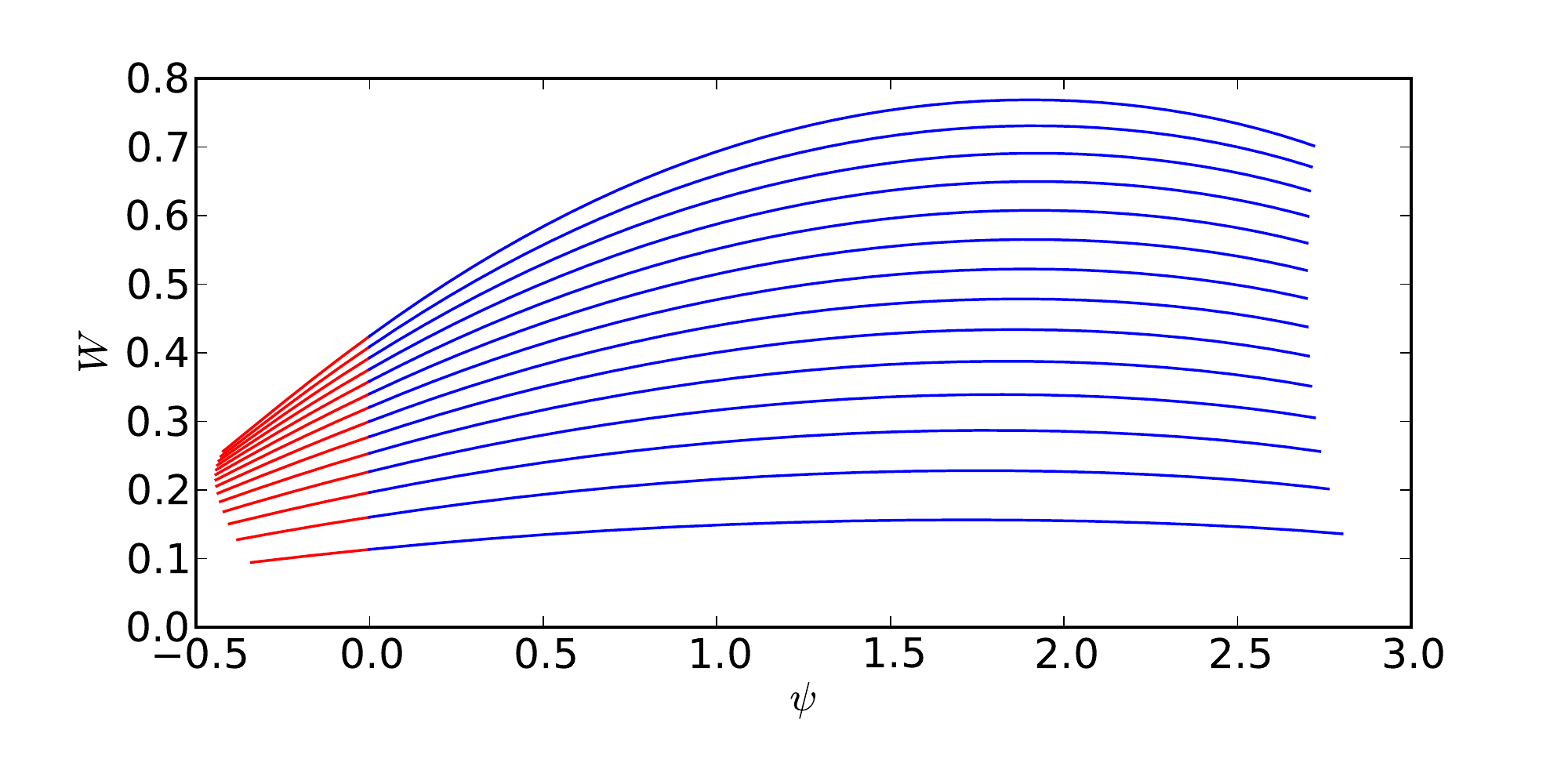}}  
\put(8,12){\includegraphics[width=8cm]{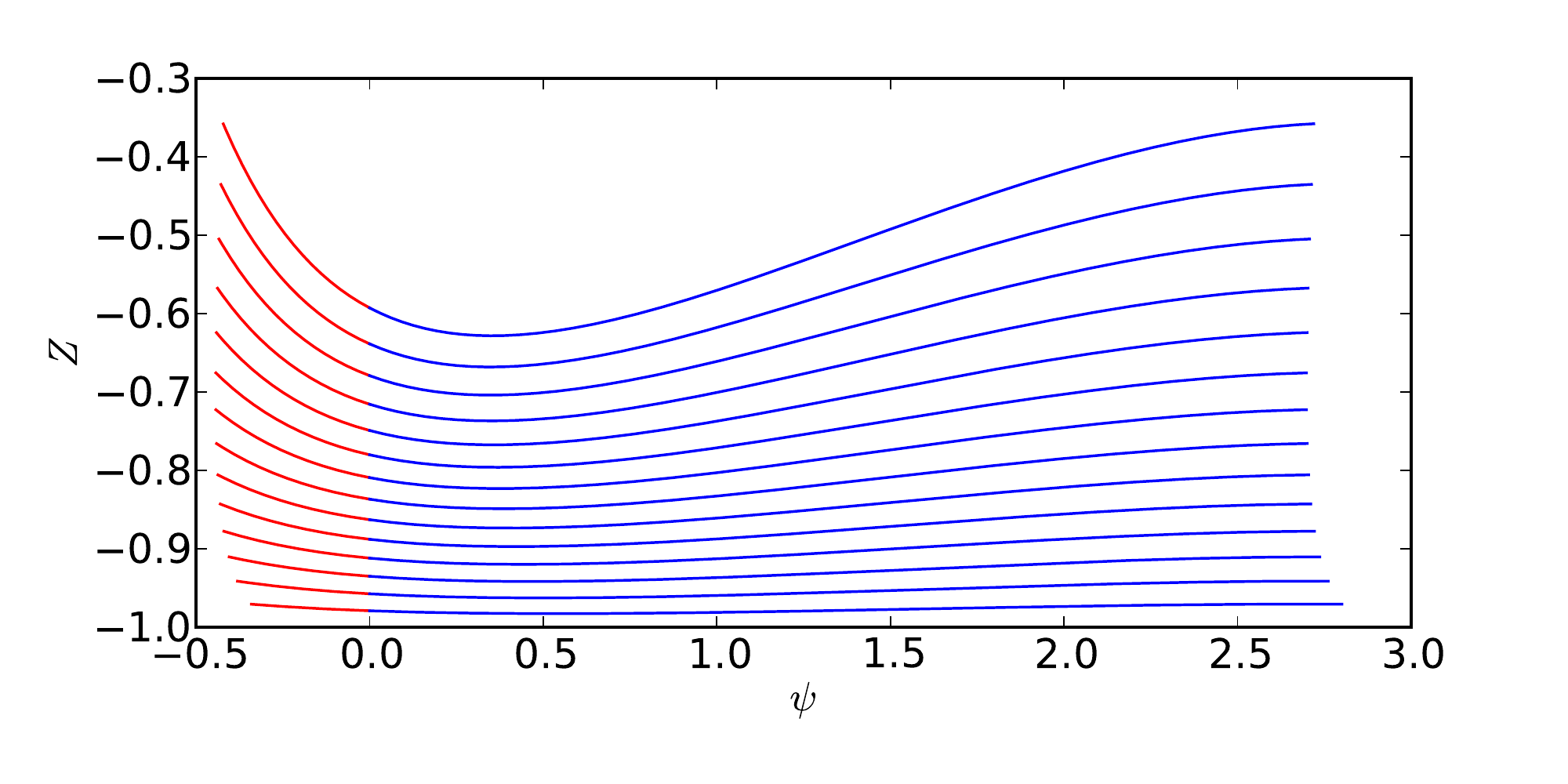}}  
\put(8,8){\includegraphics[width=8cm]{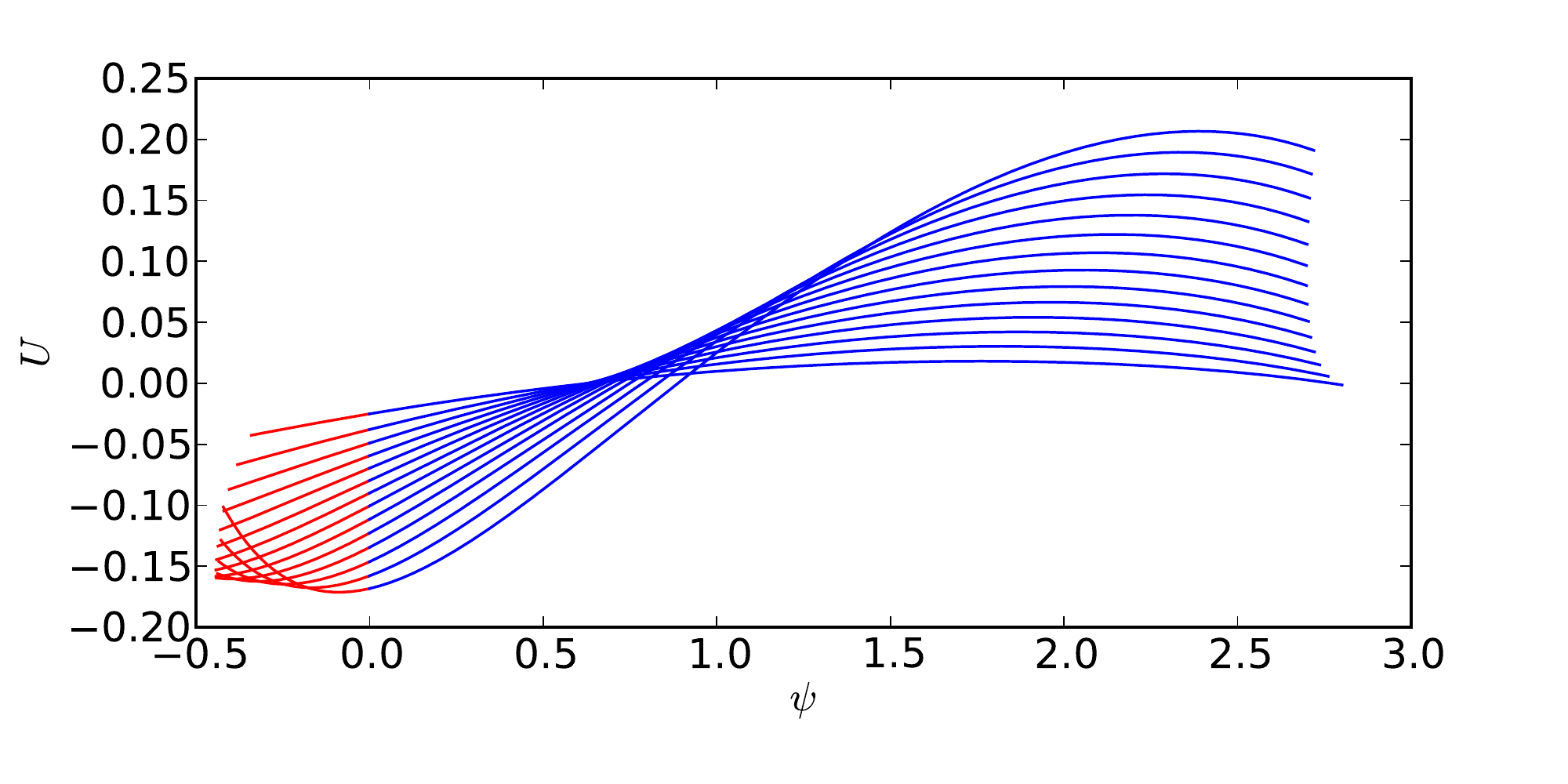}}  
\put(8,4){\includegraphics[width=8cm]{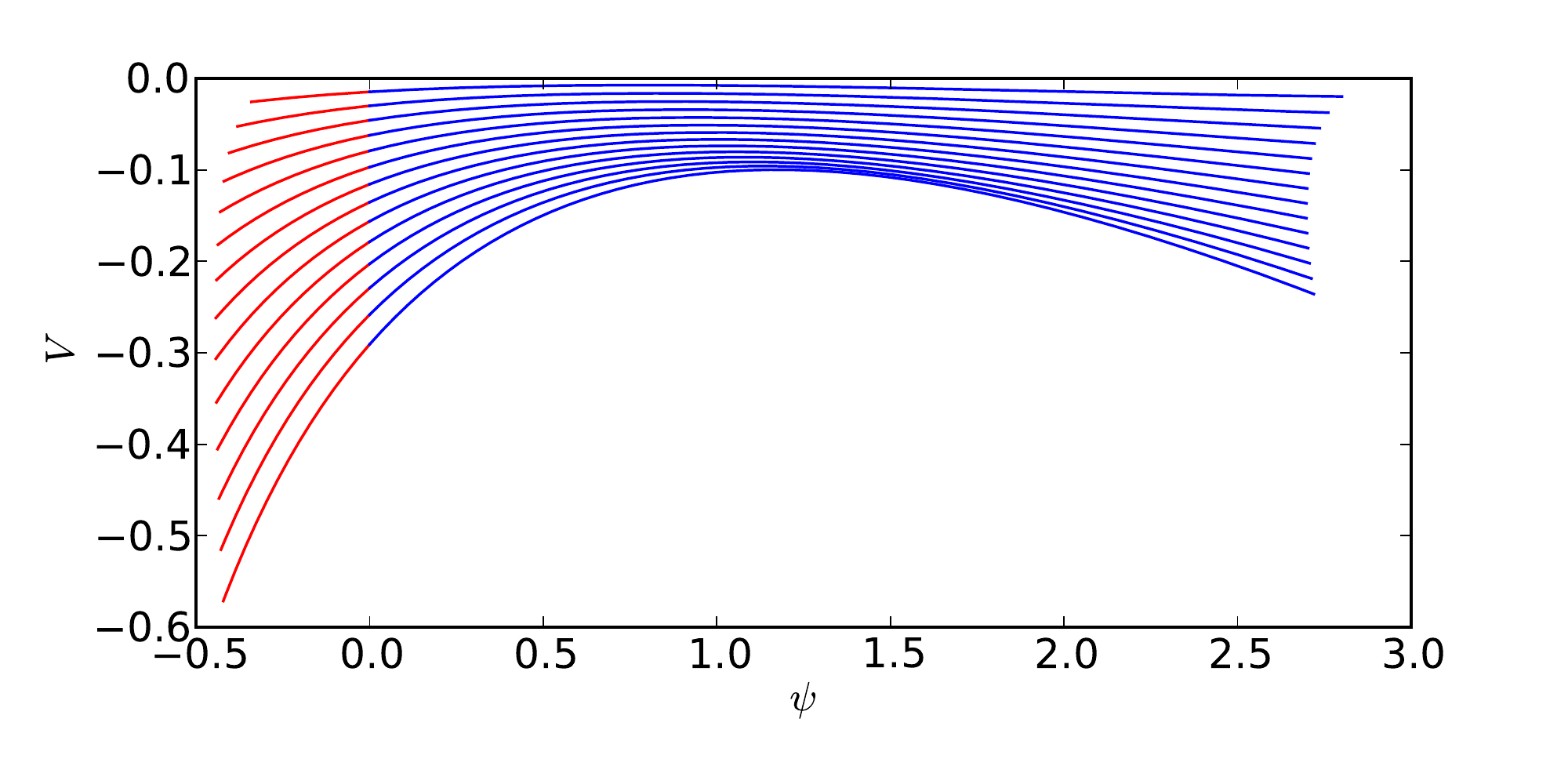}}  
\put(8,0){\includegraphics[width=8cm]{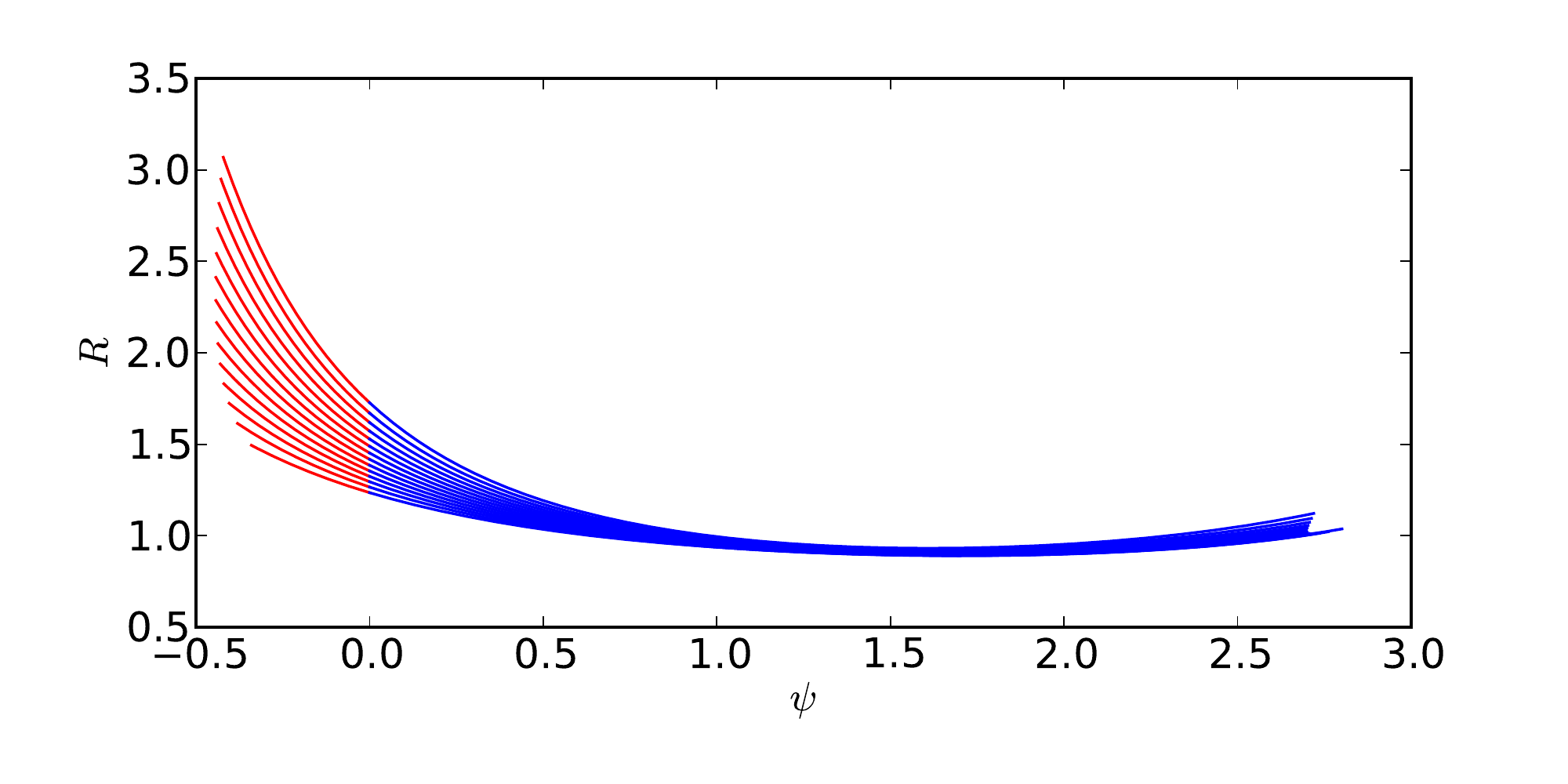}}  
\end{picture}
\caption{Mode $m=2$ for $n=1/2$ (left) and for $n=1$ (right). Various fields for a series of temperatures
equally spaced between $T_0=0.18$ (corresponding to the curves with more pronounced variations) and $T_0=0.012$
(corresponding to flatter curves).
The red parts of the curve correspond to the subsonic regions and 
the blue parts to the supersonic ones. }
\label{mode_m2}
\end{figure*}

\subsection{Numerical method}

The problem we are facing consists in solving the two ordinary
equations given by Eqs.~(\ref{ef_new_od1}-\ref{ef_new_od2}) between 0 and $2 \pi/m$. 
Mathematically, there are thus five independent parameters, the values 
$W(0)$ and $Z(0)$, as well as $B$, $T_0$ and $m$. On the other hand, there 
are two constraints coming from the shock conditions and one from the 
critical point. This implies that there are two free parameters that
should be varied. In the following the adopt the temperature, $T_0$,
and the mode number $m$ as the free parameters and we look for the
values of $B$, $W(0)$ and $Z(0)$ that satisfy the three  
constraints. Note that there might exist solutions which entail several
non-identical shocks and could constitute a broader class of
solutions that the periodic solutions considered here. 

One difficulty in solving Eqs.~(\ref{ef_new_od1}-\ref{ef_new_od2}), 
is to treat the critical point, which as described in the appendix
\ref{app:topo} is a saddle. 
To solve this system, we 
first introduce a new variable $s$ as described in the appendix \ref{app:topo}. 
The new equations Eqs.~(\ref{ef_crit_od1}-\ref{ef_crit_od3}) do not present 
any singularity and this is the ones we used to perform the numerical integration
using a standard Runge-Kutta integration.
For this purpose we first  specify a grid of the $s$ variable in decreasing order. 
To initialise the solution, it is necessary to perform an 
 expansion around the critical point as specified by Eqs.~(\ref{lin_crit1}-\ref{lin_crit3}),
in particular specifying the sign of $d \psi$. The two other quantities 
$\delta W$ and $\delta Z$ are 
along the eigenvector of the negative eigenvalue. We perform 
two integrations, one toward the left ($d \psi < 0$) and one toward the right ($d \psi > 0$). 
We then integrate toward the right (left) until either $\psi$ reach the value $2 \pi / m $ (-$2 \pi / m$)
or  reach a stagnation point, i.e. $\psi$ starts decreasing (increasing). 

Once this is done, we look for  pairs of points, which $i)$ are
located at two different sides of the critical point, $ii)$  satisfy
the Rankine-Hugoniot conditions. To do so, we define a norm, ${\cal N}$, given by:
\begin{eqnarray}
\label{norm}
\cal{N} = \\
\sqrt{ \left( {\psi_1 - \psi_2 \over {2 \pi \over m}  } -1 \right)^2 + \left({Z_1 \over Z_2} - 1 \right)^2 + \left( {W_1 W_2 \over W_c^2} - 1\right)^2 },
\nonumber
\end{eqnarray}
and we  then select the pair of points which corresponds to the minimum value. Finally, we iterate on $B$, the
spiral angle, using a simple bisection method, in order to minimise the norm.  
We use about $250,000$ grid points for each of the two trajectories
and we require to stop the iterations when 
 $B$ has varied by less than $10^{-5}$ with respect to the last
iteration. We typically obtained a clear minimum of 
$\cal{N}$ with values of the order of  $10^{-5}$.
To demonstrate that convergence has been reached we have also used 
$50,000$ grid points instead. The corresponding norm is, as expected, larger  with 
values of the order of $10^{-4}$. The solutions obtained with these
two numbers of grid points are nearly indistinguishable apart for one
particular quantity that we discuss in section~\ref{result}. 

Finally, to get a fully analytical expression of the solutions, we present in the appendix \ref{app:approx}
an approximated resolution valid at low temperature and high $m$.

\section{Results}
\label{result}

\subsection{A sample of solutions}

Figure~\ref{mode_m2} shows the $m=2$ solutions for 15 temperature
values equally spaced between 0.18 and 0.012
(the smallest temperatures correspond to the more uniform profiles). 
The left column corresponds to $n=1/2$ and the right column to $n=1$.
The red part of the curves corresponds to the subsonic regions while
the blue part represents the supersonic part of the flow. 
The critical point is at the junction of the two. 
For  large temperatures, all fields vary substantially with $\psi$
implying rather dynamical regimes.  
For example the velocity perpendicular to the spiral pattern, $W$,
becomes up to two times larger than the sound speed while 
the azimuthal velocity, $v$, 
is as small as $-0.7$ implying that the gas is then rotating at a
velocity of about $\simeq 0.3$ times the 
Keplerian velocity (equal to 1 with these units). At the largest 
temperature, the density varies by a factor of about 4  
while at smaller temperatures ($T_0 \simeq$ a few $0.01$), the
variations present a much smaller amplitude.  
Typically, at these low temperatures, 
the radial velocity, $W$, varies by about 20$\%$ while the azimuthal
velocity presents even smaller variations.  
These solutions therefore describe a flow that is close to rotational
equilibrium. Interestingly, the radial velocity, $U$, changes sign
typically around $\psi \simeq 1$.  It is always negative in the
subsonic region and positive at the end of the supersonic one.  
This structure is actually necessary to insure an inward (outward)
flow of matter (momentum) and a vanishing 
flow of angular momentum (mass), a point that will be further
discussed and quantified in Sect.~\ref{stress_sec}. 
The shape of the two families of solutions ($n=1/2$ and $n=1$) remains 
altogether very similar. As discussed previously 
and confirmed below, the solutions are however quite different in
terms of global mass and angular momentum fluxes.

In Appendix \ref{app:mode5}, we also show the $m=5$ modes.

\subsection{Dependence of spiral angles on $T_0$}

\setlength{\unitlength}{1cm}
\begin{figure} 
\begin{picture} (0,8)
\put(0,4){\includegraphics[width=8cm]{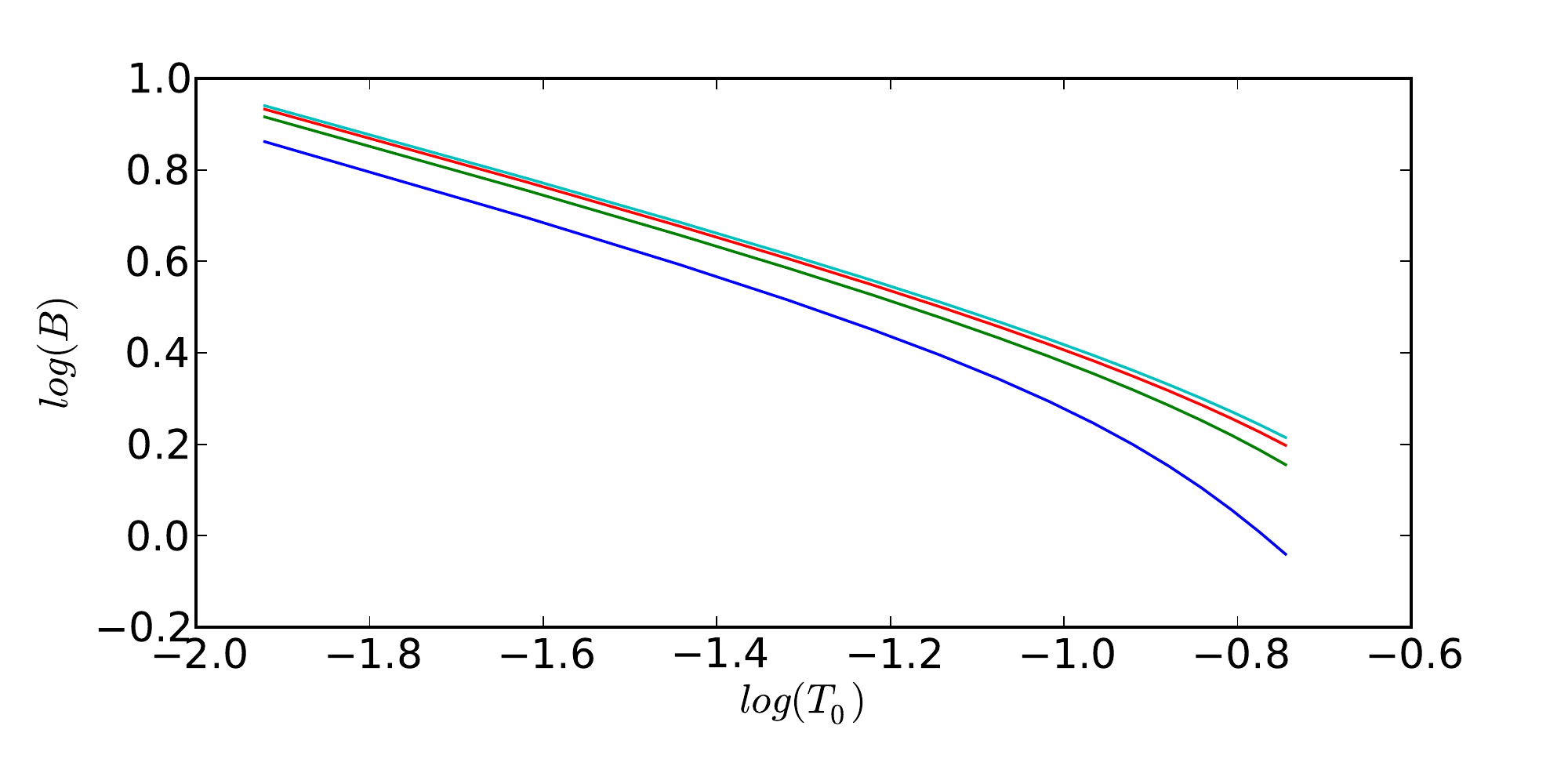}}  
\put(0,0){\includegraphics[width=8cm]{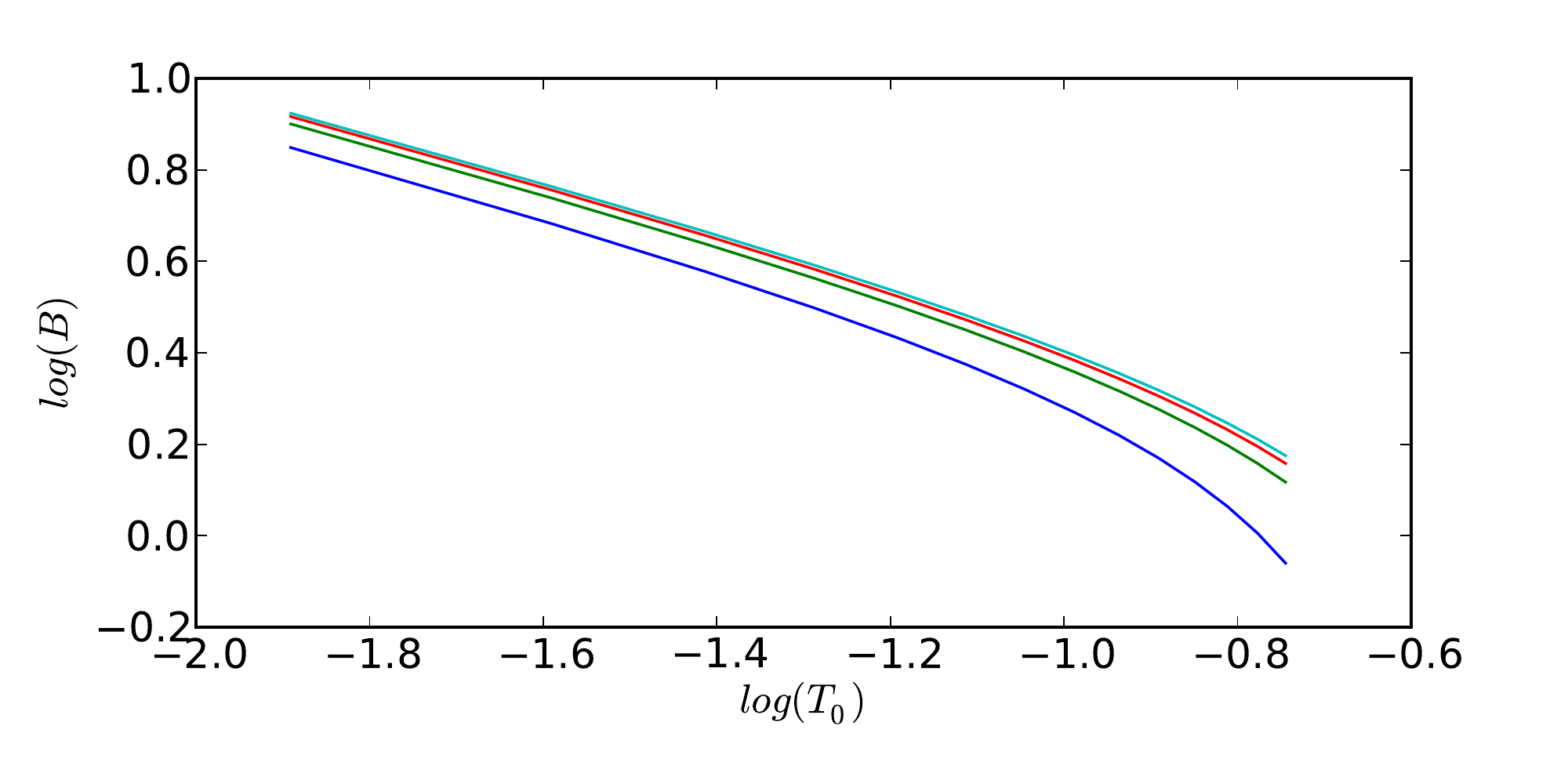}}  
\end{picture}
\caption{$B=\tan \theta$, the angle of the spiral pattern as a function of 
$T_0$ for  the four modes $m=2,3,4,5$. Top panel is for $n=1/2$ and
  bottom one for $n=1$.} 
\label{T_B}
\end{figure}


Figure~\ref{T_B} displays the logarithm of the angle of the spiral
pattern, $B=\tan \theta$, as a function of  $\log(T_0)$ for the $m=2-5$ modes. At low temperatures, $T_0 < 0.1 $, 
we find that $B \simeq T_0^{-1/2}$, meaning that the spiral pattern angle is inversely proportional
to the local sound speed.  

This can be understood easily in the weak shock regime, which is relevant in the limit $T_0\ll 1$ (thin disc limit). In this limit, the shock front speed is equal to the sound speed $C_s$ and the Keplerian rotation profile is barely perturbed by the presence of a shock. For the spiral shock to be stationary,
the Keplerian velocity projected onto the normal to the shock has to be equal to the sound speed
$C_s(R)\simeq\Omega R\cos(\theta)$, which can be transformed into
\begin{equation}
B\simeq\Big(\frac{1-T_0}{T_0}\Big)^{1/2}
\end{equation}

Stiffer variations are found for higher temperatures where $B$
increases more rapidly with $T_0$, as expected from this simple linear analysis. 
Finally, the angle also  slightly increases  with the 
mode number, $m$.


\subsection{Mass flux and stress}
\label{stress_sec}



\setlength{\unitlength}{1cm}
\begin{figure} 
\begin{picture} (0,6.9)
\put(0,3.4){\includegraphics[width=8cm]{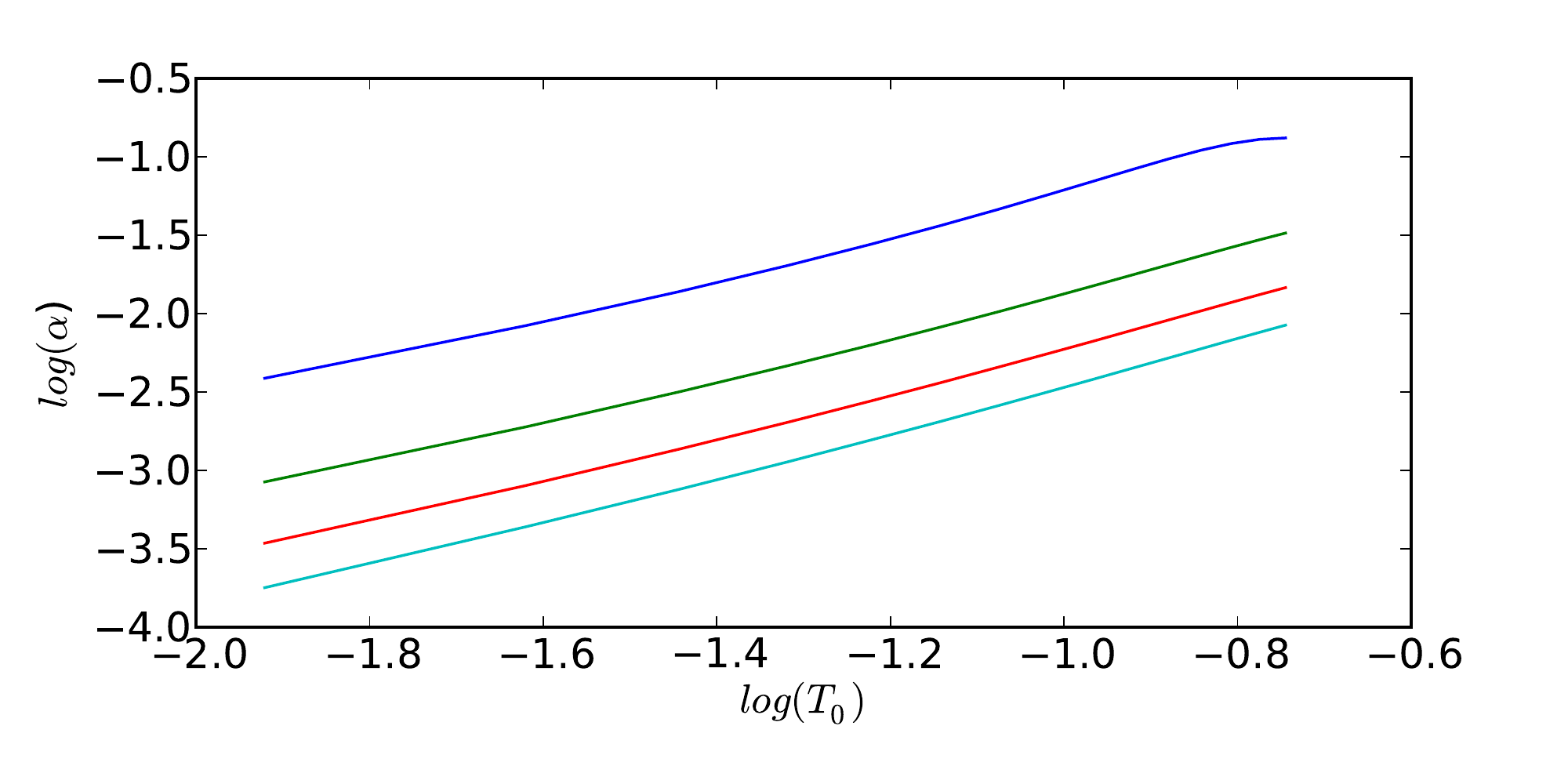}}  
\put(0,0){\includegraphics[width=8cm]{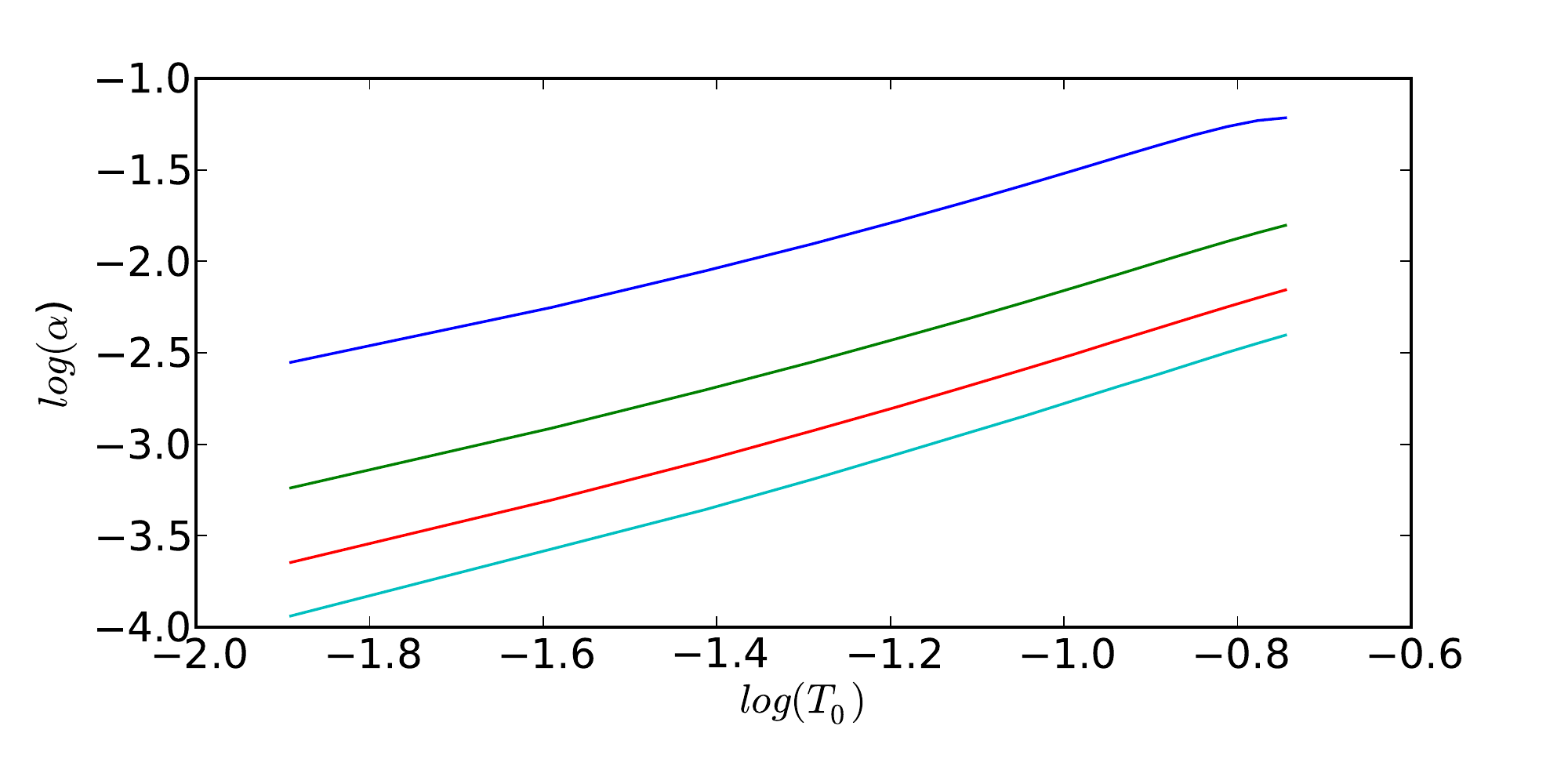}}  
\end{picture}
\caption{The $\alpha$ value for $n=1/2$ (top panel)
and $n=1$ (bottom panel)
as a function of 
$T_0$ for  the four modes $m=2,3,4,5$. }

\label{flux_mass}
\end{figure}


\subsubsection{Definitions}

We now describe and quantify the global mass and momentum fluxes
associated with these solutions. More precisely, we are interested in
the flux of mass $\int u_r \Sigma r d\phi$ that we write as
\begin{eqnarray}
\label{mass_flux}
{\cal F} = \int _0 ^{2 \pi} R(\psi) U(\psi) d \psi, 
\end{eqnarray}
It is known \citep[e.g.][]{balbpap99} that the fluxes of mass and the stress are related to each other through the 
relation 
\begin{eqnarray}
\label{stress_eq}
< \Sigma u_r  > = -{1 \over \partial _r (\Omega r^2) r  } \partial_r  < \Sigma r^2 \delta u_r \delta  u_\theta > .
\end{eqnarray}
In this expression, $\delta u_r= u_r - <u_r>$,  $\delta u_\theta=
u_\theta - <u_\theta>$ and $\Omega$ is the mean rotation value,
$\Omega = <u_\theta> / r$. While this relation is a good approximation
in the general case (because it neglects a time-dependent term), we
stress here that it is an exact relation in the present, stationary
case. It must therefore be satisfied and constitutes a test for the
accuracy of the numerical solutions.

With the self-similar variables, 
Eq.~(\ref{stress_eq}) becomes
\begin{eqnarray}
\label{stress_simil}
{\cal F} = {\cal S} = -{2(1-n) \over  1 + {\cal V}  }  \int _0 ^{2 \pi} R(\psi) U(\psi) (V(\psi)-{\cal V}) d \psi.
\end{eqnarray}
where
\begin{eqnarray}
{\cal V} = { \int _0 ^{2 \pi} R(\psi) V(\psi) d \psi \over \int _0 ^{2 \pi} R(\psi)  d \psi} , 
\end{eqnarray}
is the mean value of $V$.
While for $n=1/2$ the coefficient $2(1-n)$ is equal to 1, it is equal to 0 for $n=1$, which indicates 
that in this latter case the mass flux should be zero as already discussed.

It is usual to define the quantity $\alpha$ as given by 
\begin{eqnarray}
\label{alpha}
\alpha = {< \Sigma \delta u_r \delta  u_\theta > \over < \Sigma C_s^2 >}, 
\end{eqnarray}
which leads to 
\begin{eqnarray}
\label{alpha_ss}
\alpha = {\int _0 ^{2 \pi}  R(\psi) U(\psi) (V(\psi)-{\cal V}) d \psi  \over T_0 \int _0 ^{2 \pi}  R(\psi)  d \psi}.  
\end{eqnarray}
In this last expression $U$ is used instead of $U - <U>$ since $<V - {\cal V}>=0$.

Finally, we also compute the flux of angular momentum through the disc, which as discussed before is expected 
to vanish for $n=1/2$
\begin{eqnarray}
\label{mom_flux}
{\cal F}_{mom} = \int _0 ^{2 \pi} R(\psi) U(\psi) (1 + V(\psi) ) d \psi. 
\end{eqnarray}
Note that when $n=1$, since the mass flux vanishes, $\int R(\psi) U(\psi) d \psi =0$, we
have the identity
\begin{eqnarray}
\label{mom_flux2}
{\cal F}_{mom} = {\cal F}_{mom, \, n=1} = \alpha T_0 \int _0 ^{2 \pi} R(\psi) d \psi. 
\end{eqnarray}
Note that this expression  is valid only if the flux of mass vanishes.

In the following section, since all quantities depend on $\alpha$, we restrict our attention 
to its value.

\subsubsection{Resulting fluxes: the $\alpha$ value}

Top pannel of Fig.~\ref{flux_mass} displays the values of $\alpha$
as a function of $T_0$ for the modes $m=2,3,4,5$. First of all, we see that 
significant stresses leading to significant  mass fluxes  are
inferred. In terms of the canonical 
$\alpha$, values as high as $\simeq$ 0.1 are obtained at large
temperature. We also find that $\alpha$ scales 
with temperature roughly as $\alpha \propto T_0^{3/2}$ and decreases
as $m$ increases. This is  expected since $\alpha$ is
proportional to the product of the velocity fluctuations. Indeed, the
velocity fields vary over a smaller domain and the typical value of
the gradient is obtained at the critical point and does not vary
significantly with $m$ (since it depends only on $B$ which does not
vary strongly with $m$). Physically, higher $m$ modes tends to be
closer to an axisymmetric configuration.


Note that we have verified that the flux of mass obtained from expression ${\cal F}$ and ${\cal S}$ 
are very close to each other. There are however not identical because of the numerical integration. 
In particular, the flux of mass calculated using expression ${\cal F}$ appears to be more noisy. 
This is because $U$ changes sign and the integral values of the 
regions where it is either positive or negative are very close. On the
other hand, when $\alpha$ is evaluated, $V - {\cal V}$ is also
changing sign at the same locations as $U$ (reflecting the fact that
the radial and azimuthal velocity components 
are highly correlated) and therefore the sign of the integrand is
generally positive, making it less sensitive to the fluctuations.

We have also  verified that as expected, the flux of angular momentum is extremely small and indeed equal to zero within
the accuracy of the calculation (not displayed here for conciseness). 

\setlength{\unitlength}{1cm}
\begin{figure} 
\begin{picture} (0,7)
\put(0,0){\includegraphics[width=8cm]{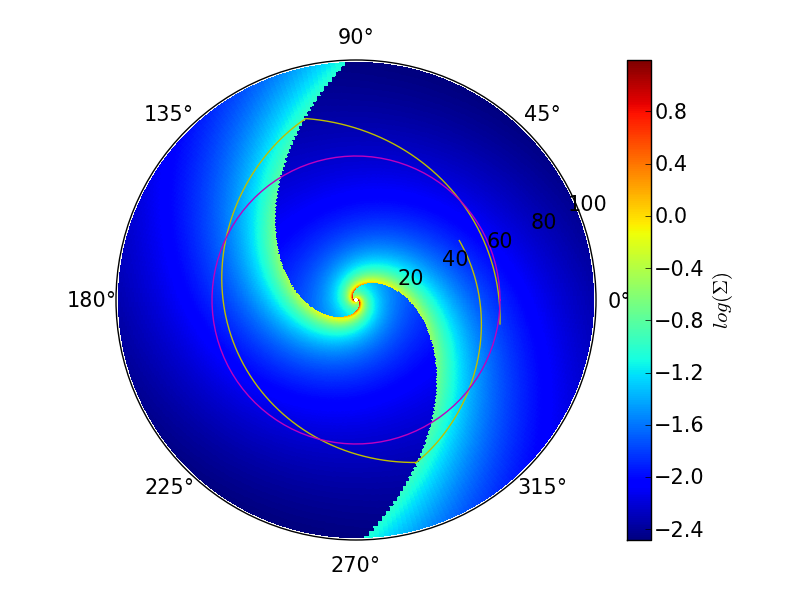}}  
\end{picture}
\caption{Case $n=1/2$. Bidimensional representation of the
  self-similar spiral pattern for  
a temperature namely $T_0=$0.17  leading to values of 
$B=\tan \theta$ equal to 1. and to a value of 
$\theta$ equal to about 45 degrees. The yellow
line corresponds to the trajectory of a fluid particle and the red circle
shows a circular orbit that the same fluid particle would have in a
symmetrical disc. Fluid particles are found to spiral inward because
of dissipation.}
\label{spiral_picture}
\end{figure}

\setlength{\unitlength}{1cm}
\begin{figure} 
\begin{picture} (0,4)
\put(0,0){\includegraphics[width=8cm]{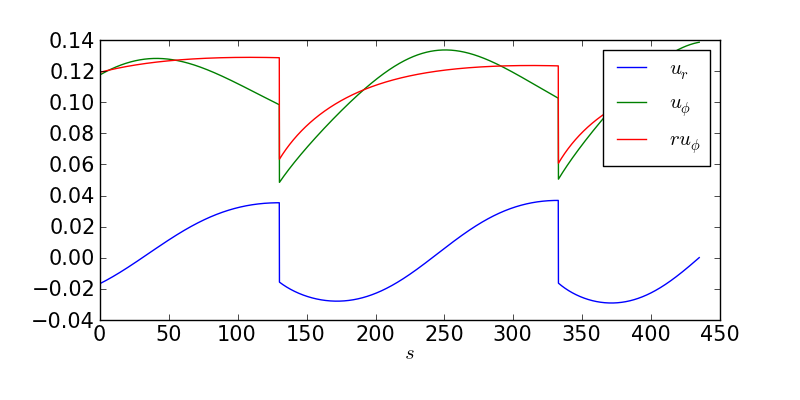}}  
\end{picture}
\caption{Radial and azimuthal velocities and angular momentum of the
  fluid particles along the trajectories displayed in
  Fig.~\ref{spiral_picture} corresponding to a radius of $r \simeq
60$ for $n=1/2$.}
\label{spiral_traj}
\end{figure}

\setlength{\unitlength}{1cm}
\begin{figure} 
\begin{picture} (0,7)
\put(0,0){\includegraphics[width=8cm]{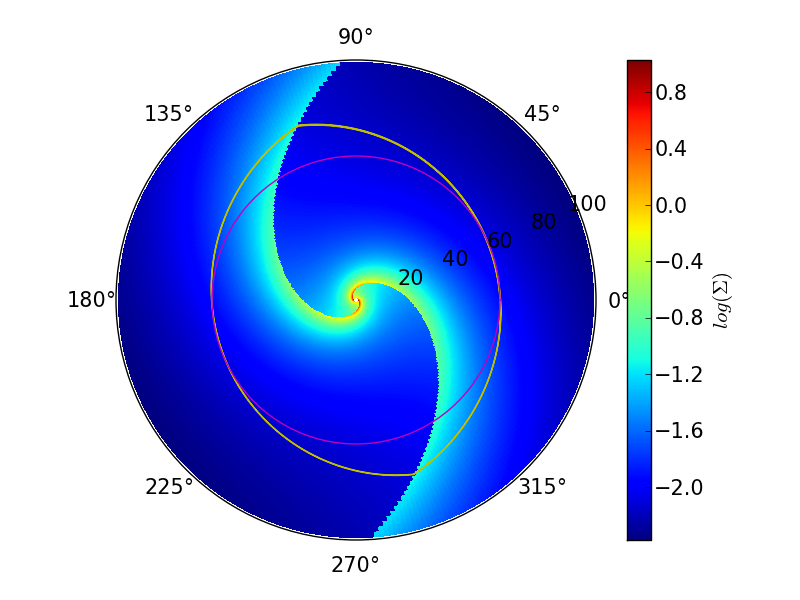}}  
\end{picture}
\caption{Case $n=1$. Bidimensional representation of the self-similar spiral pattern for 
  $T_0=$0.17, leading to values of 
$B=\tan \theta$ equal to $\simeq$1.0 and
$\theta$ equal to about 45 degrees. The yellow line corresponds 
to the trajectory of a fluid particle and the red circle shows a circular orbit that the 
same fluid particle would have in a symmetrical disc. The trajectory of the fluid particle 
shows that while the fluid particle has a non circular orbit, it is nevertheless closed. This is because the mass
flux vanishes. There is however an outwards flux of angular momentum. }
\label{spiral_picture_n_1}
\end{figure}

Figure~\ref{flux_mass} also shows (bottom panel) $\alpha$
 for $n=1$ as a function of $T_0$. 
The value of $\alpha$ remains quite similar to the case $n=1/2$ although 
a little lower (by typically about 20$\%$).  This confirms that the influence of 
the density profile onto the various fields is not too drastic.
The fluxes are however, as already discussed, quite different. The mass flux 
is equal to zero within numerical precision (not displayed here for
conciseness), while the momentum flux is  
a simple power law of the temperature, $\simeq T_0^{-2.5}$ similar to the dependence 
of the mass flux when $n=1/2$.


\subsection{Lagrangian analysis}

In order to gain physical insight, it is interesting to  follow the trajectory of a 
fluid particle, that is to say a particle that follows the stream lines. 
Figure~\ref{spiral_picture} displays a bidimensional view of the density 
field for a temperature $T_0=$0.17 corresponding to the  angle $\theta=$45 degrees. 
The yellow curves show the trajectory of a fluid particle. It was
obtained by simply solving the two equations
\begin{eqnarray} 
{d x \over dt} = v_x \, , \\
{d y \over dt} = v_y \, .
\end{eqnarray} 
A circle representing the trajectory that would be followed by a fluid
particle in a symmetrical unperturbed disc is also plotted for comparison. 
Because of the self-similar nature of the solutions, we stress that all
trajectories are identical to the one displayed there once rescaled
and rotated.  

As can be seen on figure~\ref{spiral_picture}, when the fluid particle encounters the shock, it is deflected inwards (since the 
velocity, $v_\perp$ decreases while $v_\parallel$ is unchanged). Consequently, it tends to fall toward
the disc center. However, as  is clear from Fig.~\ref{spiral_picture}, there is a 
density and therefore a pressure gradient, due to the spiral structure, that is pushing the 
fluid particle outwards. Therefore the fluid particle is decelerated
in the radial direction and accelerated in the azimuthal direction.
This is even clearer in Fig.~\ref{spiral_traj} in which the radial and
azimuthal velocity along with the angular momentum are displayed along
the fluid particle trajectory as a function of the curvilinear  
abscissa $s$: $u_r$ increases continuously after the shock up to the next shock
(with a short phase during which it further decreases for the largest $T_0$). 
Similarly, $u_\phi$ increases continuously 
  between $s=30$ and $s=130$ and
then decreases. This last phase is simply due to the fact that since
the fluid particle is moving outwards, its Keplerian velocity
decreases. The evolution of the specific angular momentum is also
enlightening. After a steep decrease through  
the shock, it increases continuously and tends toward a constant value. The physical picture is 
thus that at the shock, the fluid particle is suddenly slowed down and this results in an exchange of 
angular momentum, through the pressure forces with the post shock gas. After this point, the particle's
momentum increases because of the pressure gradient. Thus the fluid particle is being given 
angular momentum from the gas that is located at smaller radii. This momentum is then carried 
along up to the next shock when it will be delivered to higher radii material. 

Although the radii of the fluid particle varies non-monotonically
during one cycle (i.e. during 2 shocks), it is globally decreasing
with time and the particles spiral inwards as expected (the mass flux
being negative in that case).

By comparison, Fig.~\ref{spiral_picture_n_1} shows the density profile
and fluid particle trajectory for $n=1$ and $B \simeq 1$ (similar to the first panel
of Fig.~\ref{spiral_picture}).  As expected, the orbit
(yellow curve) is closed even though it is not circular. 

\section{Conclusion}

We have investigated self-similar solutions of a spiral pattern within a disc. 
They are similar to the one studied by \citet{spruit87} but a
different assumption is made regarding the temperature
distribution. In addition, different density profiles are considered. 
These solutions, which are self-similar in radius, depends on the azimuthal angle,  
and describe a non-linear spiral wave propagating in a centrifugally supported and 
locally isothermal disc. They feature shocks at which location dissipation
takes place. Because the flow is supersonic when the gas enters the
shock and subsonic as it emerges, the solutions present a critical
sonic point, which describes the transition from subsonic to
supersonic motions. Since the equations have eventually to be solved numerically, 
we have carefully studied the nature of this critical
point and have shown that almost everywhere it is a saddle rather than a node. 

Numerically solving the ordinary equations under the constraint that
the flow must satisfy the Rankine-Hugoniot conditions through the
shock, we obtained a series of profiles for various temperatures and 
mode number $m$. We inferred the values of $\alpha$ and showed that it
can be as large as $\sim 0.1$ for 
the thickest discs for which solutions exist ($h/r \simeq 1/3$). For smaller temperatures, 
it then drops as $T^{1.5}$ or equivalently as $(h/r)^3$.  We found that the spiral angle, $\theta$, increases 
when $T$  diminishes roughly as $\theta = \arctan (r/h)$. 
Two density profiles are being explored. For the first one ($n=1/2$), we find a non-vanishing mass flux 
and a zero angular momentum flux. For the second one ($n=1$), the first vanishes but not the latter.
The parameter  $\alpha$ is however very similar for these two cases, with the steeper profile 
presenting slightly lower values. 

From a Lagrangian analysis of the solutions, it is concluded that in the $n=1/2$ case the fluid particles 
spiral inwards and undergo a series of shocks, followed by a pressure acceleration
due to the global spiral pattern. During these two phases the fluid particles are 
respectively losing and gaining angular momentum due to momentum exchange with the 
surrounding gas. This leads to an inward flux of mass through the disc.
In the $n=1$ case, the fluid particles follow a non-circular closed orbit. There is however an 
outward flux of angular momentum.
While these two types of solutions present different behaviours in terms of fluxes, they are rather similar 
and typically differ by only a few tens of percents. Their Lagrangian behaviours are also very similar.

Although restricted to particular temperature and density profiles, the existence of these solutions
suggests that external perturbations exerted onto accretion discs can propagate deep into the discs
and therefore should not be ignored. In many systems, the most natural source of external perturbations is 
the accretion of external gas which produce shocks at the disc surface.

\section*{ACKNOWLEDGMENTS}
We thank the anonymous referee for a constructive and helpful report. 
This research has received funding from the European Research Council under the European
 Community's Seventh Framework Programme (FP7/2007-2013 Grant Agreement no. 306483).

\appendix
\section{\label{app:topo}Topology of the critical points}
The topological nature of the critical point is worth studying as it also 
constraints the values of $T_0$ and $B$. In particular, it is 
important to know whether it is a node or a saddle. In the first case, 
a  one dimensional ensemble of  trajectories will be able to cross it while 
in the second case, only a discrete set of trajectories will have to be considered.
To achieve this we introduce a new variable, $s$, such that

\begin{eqnarray}
\label{ef_crit_od1}
     {d \psi \over ds }  =  {T_0 -  W^2},   \\
\label{ef_crit_od2}
\nonumber
      {d W \over ds }  = -{W \over 2 (1+B^2) }   &&   \times  \\
\nonumber
    \left( B W^2 + 2 B Z ^2 \right. - WZ &+& \left.  B \left( 2 (n+1) T_0-2 \right) \right. \\
 &-& \left. T_0 \left(2n - 1  \right)  {B W + Z \over W} \right),  \\
\label{ef_crit_od3}
      {d Z \over ds }  =  {T_0 - W^2 \over  (1+B^2) W} && \times  \\
\nonumber
\left( W^2 + {Z^2 \over 2} - {B \over 2} W Z  \right. &+&  \left( (n+1) T_0 -1 \right) \bigg).
\end{eqnarray}
Note that it is also possible to consider $-s$ instead of $s$, however this leads to 
unphysical solutions that entail a rarefaction shock, i.e. the gas enters the shock 
subsonically and leaves it supersonically.

To study the topology of the critical point, we make an expansion in its neighbourhood 
and obtain a linear system
\begin{eqnarray}
\label{lin_crit1}
     {d \delta \psi \over ds }  &=&  - 2  W_c \delta W,   \\
\label{lin_crit2}
      {d \delta W \over ds }  &=&    M_{WW} \delta W + M_{WZ} \delta Z =  \\
\nonumber
{ - W_c \over 2 (1+B^2) } &\times&
    \left( \left( 2 B W_c - Z_c + T_0 \left(2 n- 1  \right) {Z_c \over W_c^2} \right) \delta W \right. \\
\nonumber
 &+& \left. \left(4 B Z_c - W_c - T_0 \left( 2 n - 1  \right) {1 \over W_c} \right) \delta Z   \right),  \\
\label{lin_crit3}
      {d \delta Z \over ds }  &=&  M_{ZW} \delta W = \\
\nonumber
{ - 2 \delta W \over  (1+B^2) } &\times&
 \left( W_c^2 + {Z_c^2 \over 2} \right. - \left. {B \over 2} W_c Z_c + \left( (n+1) T_0 -1 \right) \right).   
\end{eqnarray}
The matrix of this  linear system admits three eigenvalues, 0, and 
\begin{eqnarray}
\lambda_{\pm} = {1 \over 2} \left( M_{WW} \pm \sqrt{ M_{WW} ^2 + 4 M_{WZ} M_{ZW}} \right).
\end{eqnarray}
Let $Y_i^0$ be the three eigenvectors associated to the three eigenvalues. 
In the neighborhood of the critical point, the solutions of the linear system 
Eqs.~(\ref{lin_crit1}-\ref{lin_crit3}) are linear combination of
$Y_i^0$ and can be written as
\begin{eqnarray}
\label{exp_sol}
Y(s) = \Sigma _{i=1,3} \, \alpha_i \exp( \lambda_i s)  Y_i^0, 
\end{eqnarray}
where $\alpha_i$ are real coefficients.
Obviously, if $\lambda_i > 0$, $Y(s) \rightarrow \infty$ implying 
that $\alpha_i$ must be 0 in order for the corresponding solution 
to cross the critical point. It is therefore important to 
know the sign of $\lambda_+$ and $\lambda_-$. 
Another possibility is that $\lambda_+$ and $\lambda_-$ are complex
conjugate. In this case, the solutions approach the critical point 
with an oscillating behaviour. Such solutions are not physical 
either since they imply multi-valuate physical variables at the same
location $\psi$. It is therefore important to study the signs
 of $M_{WW}$, $M_{WZ}$, $M_{ZW}$ and $M_{WW} ^2 + 4 M_{WZ} M_{ZW}$.
Plunging the expression of $W_c$ and $Z_c$ (selecting the ``-'' sign 
in Eq.~\ref{Z_crit2}), we obtain their values. 
It is easy to verify that $M_{WW}<0$ while $M_{WZ}>0$.

\setlength{\unitlength}{1cm}
\begin{figure}[h!] 
\begin{picture} (0,4)
\put(0,0){\includegraphics[width=8cm]{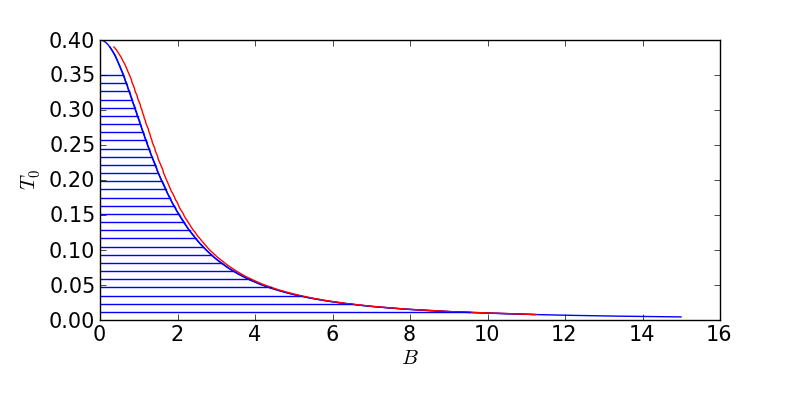}}  
\end{picture}
\caption{Topology of the critical points for $n=1/2$. Blue: points satisfying $M_{ZW}=0$.
This represents the transition from saddle to node critical points (one negative and one positive eigenvalues). 
Red: points satisfying $M_{WW} ^2 + 4 M_{WZ} M_{ZW}$. Transition between the node critical points 
to oscillatory critical points (two complex conjugate eigenvalues). 
The physical solutions are to be searched in the dashed area.}
\label{T_B_crit}
\end{figure}

The sign of $M_{ZW}$ and $M_{WW} ^2 + 4 M_{WZ} M_{ZW}$ is less straightforward and 
we have studied their values numerically (the sign of $M_{ZW}$ can be obtained through 
a second order polynomial). Figure~\ref{T_B_crit} shows the curves in the $B-T_0$ plane, 
which correspond to $M_{ZW}=0$ (blue curve) and $M_{WW} ^2 + 4 M_{WZ} M_{ZW}=0$ (red curve).
The possible solutions are located in the dashed region where there is one 
 negative eigenvalue. In this region the critical points are saddle. 
Strictly speaking, it is also possible to have solutions with critical points
located in between the blue and the red curves where the eigenvalues are both negative 
and the critical points are nodes. These solution however would present a weak discontinuity, 
that is to say the derivatives of the fluid variables (density and velocity) are discontinuous.

\section{\label{app:mode5} The $m=5$ mode}

\setlength{\unitlength}{1cm}
\begin{figure*} 
\begin{picture} (0,20)
\put(0,16){\includegraphics[width=8cm]{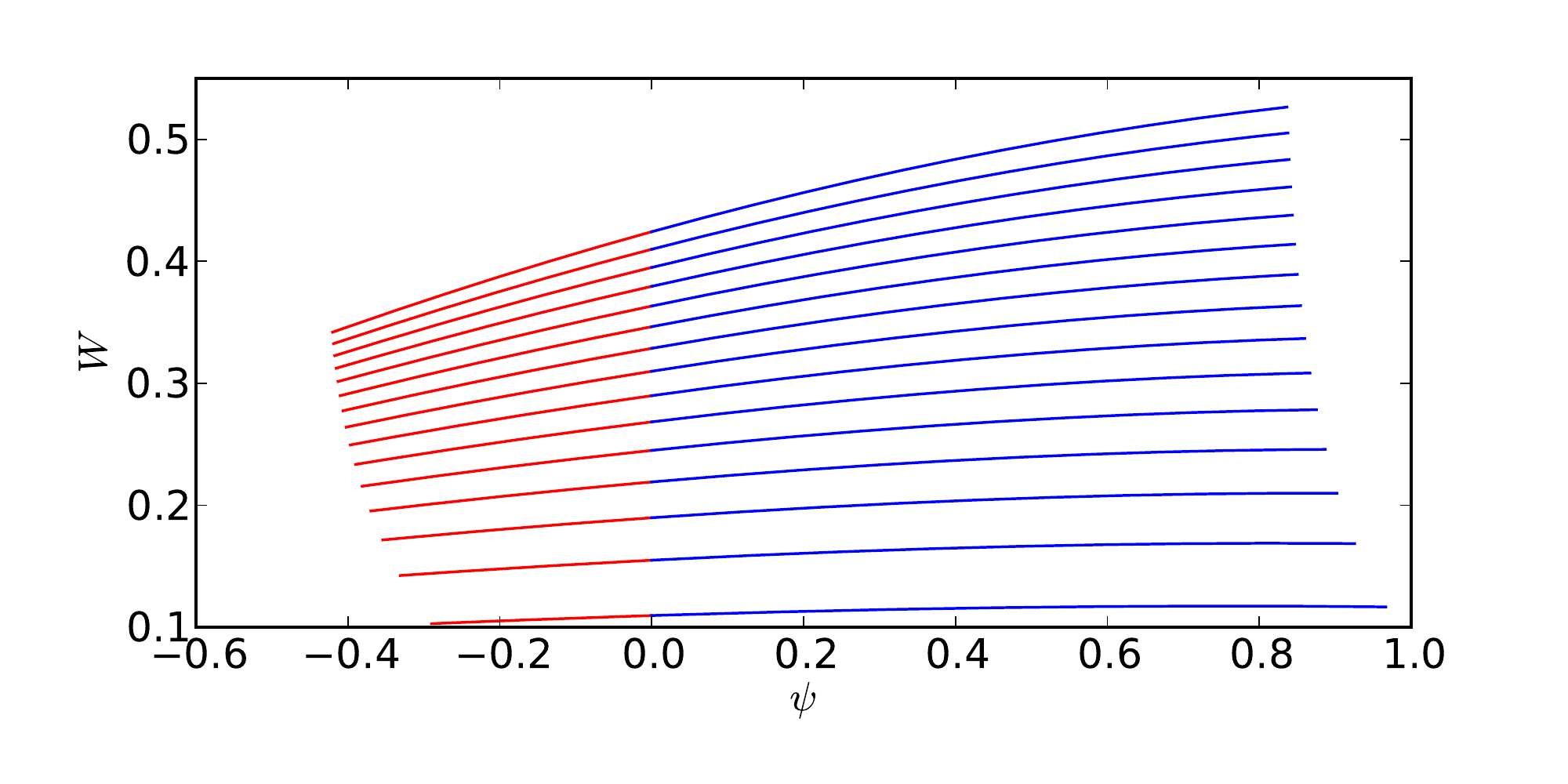}}  
\put(0,12){\includegraphics[width=8cm]{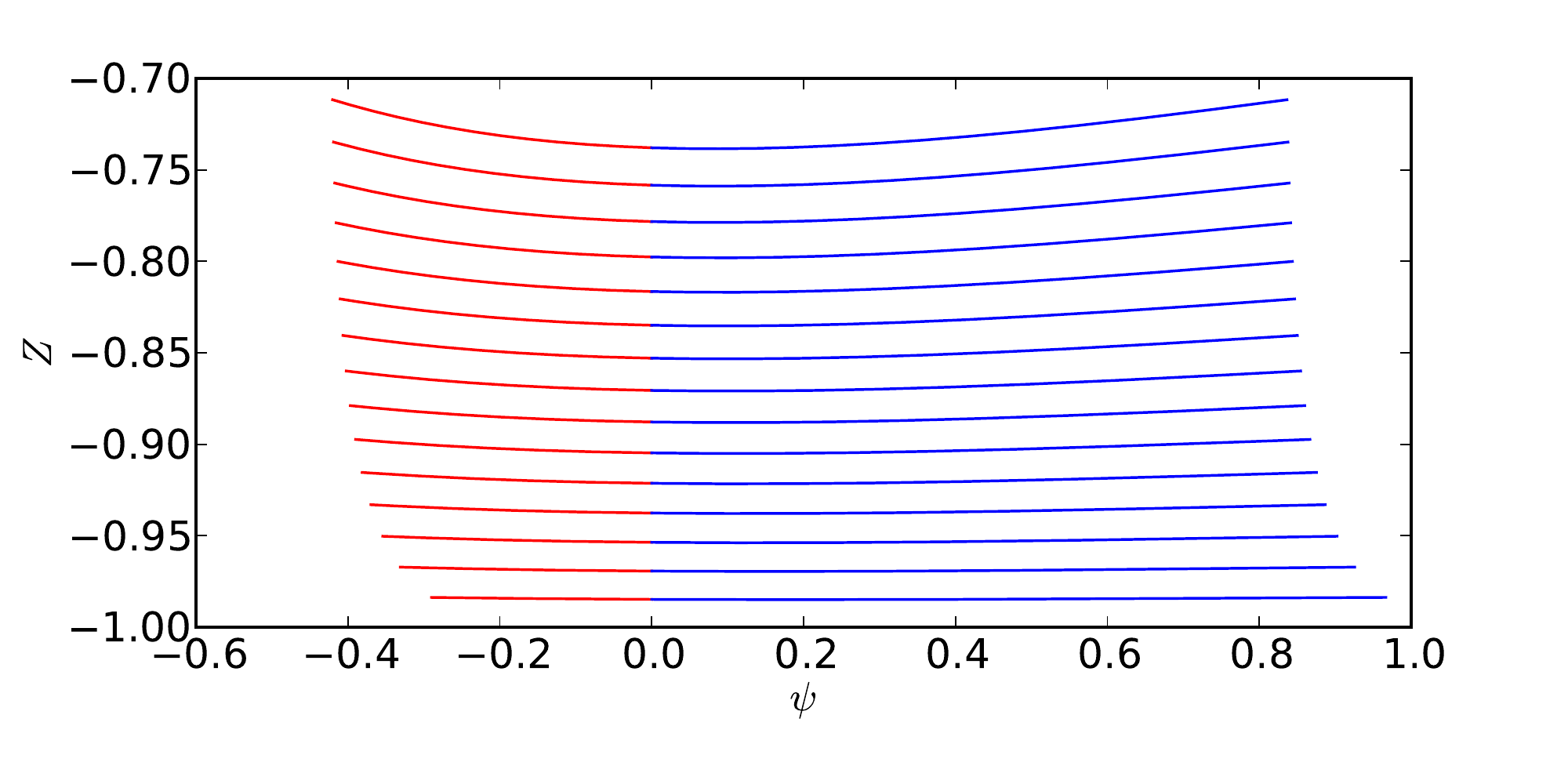}}  
\put(0,8){\includegraphics[width=8cm]{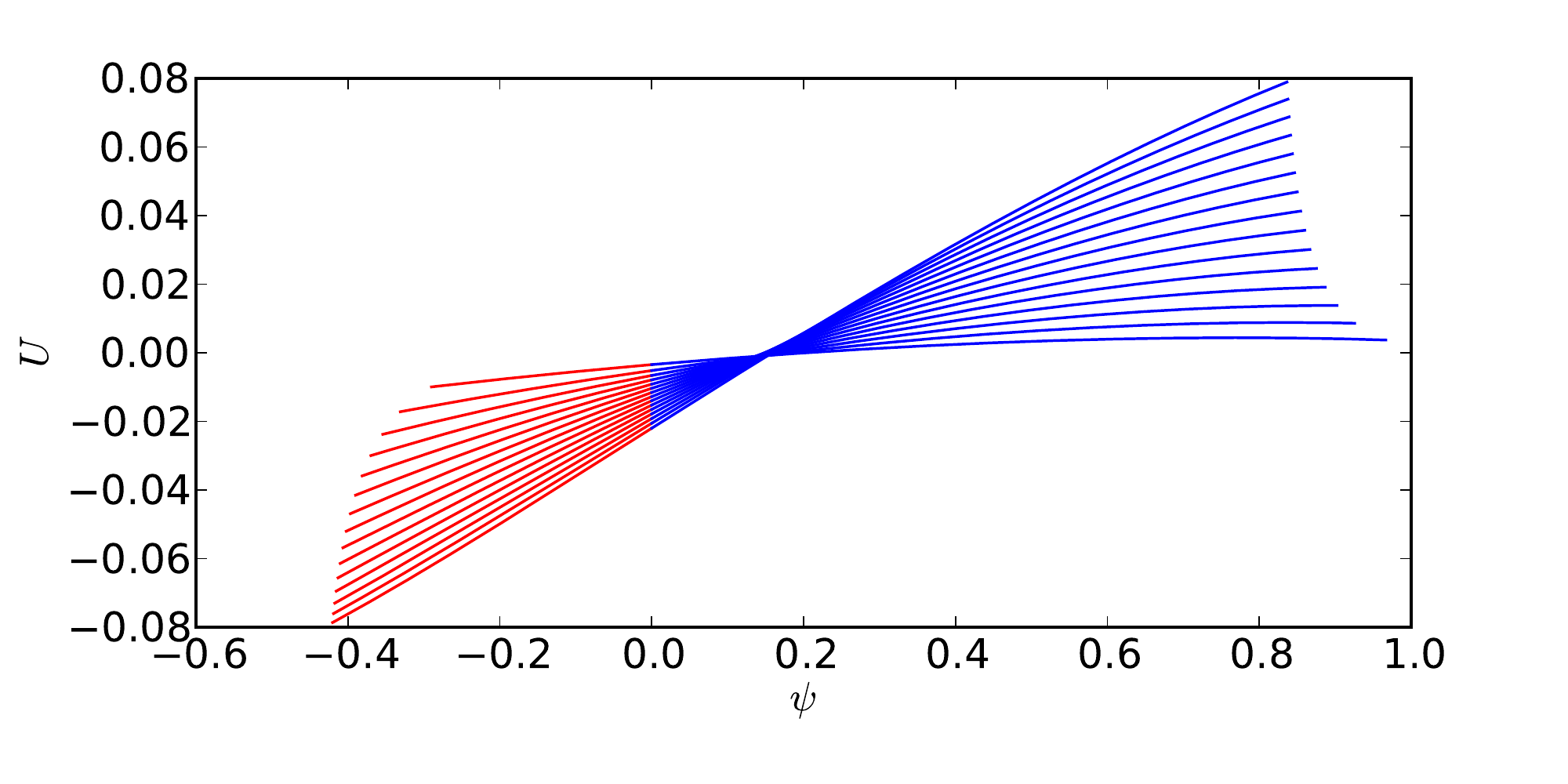}}  
\put(0,4){\includegraphics[width=8cm]{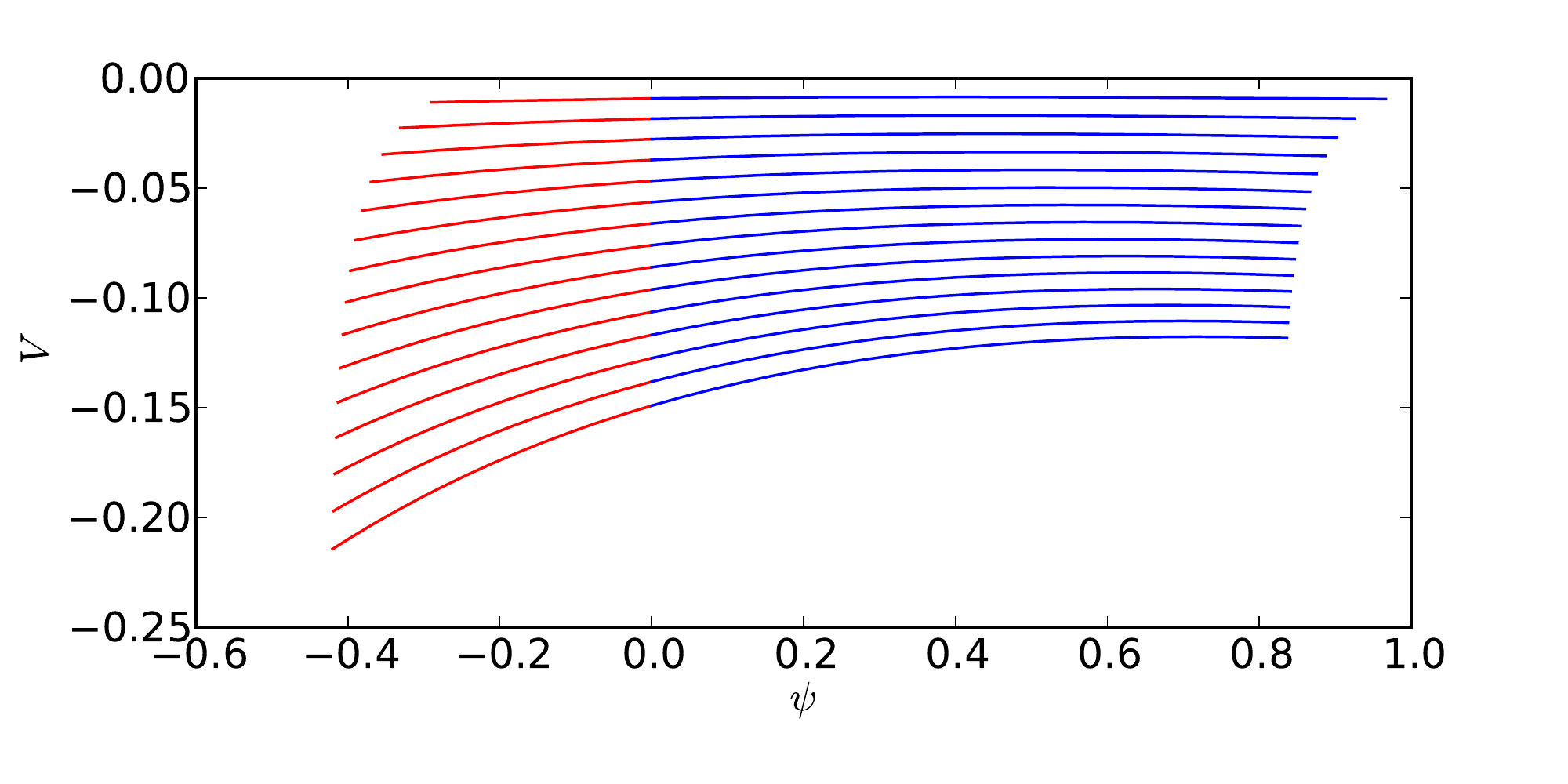}}  
\put(0,0){\includegraphics[width=8cm]{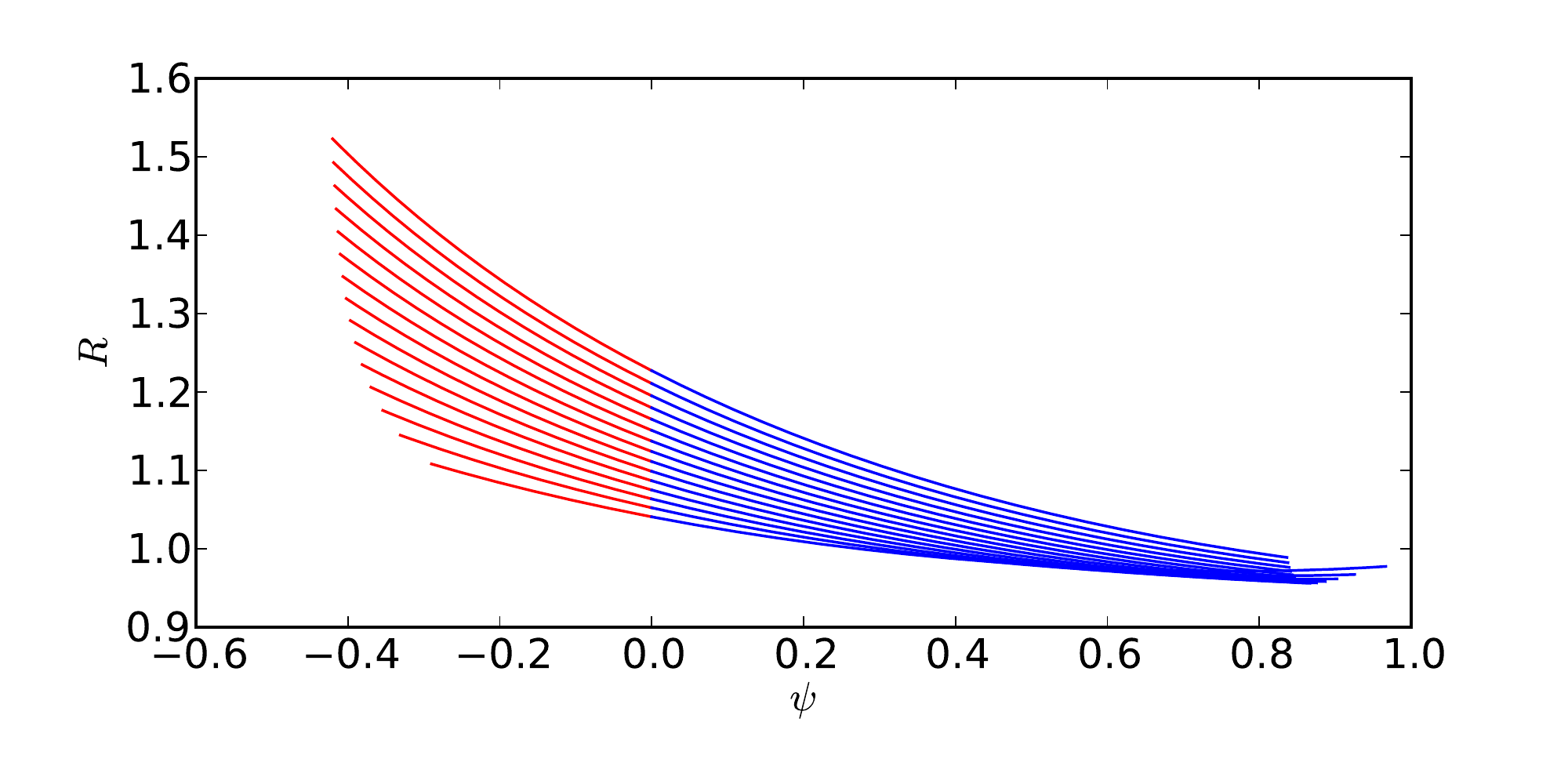}}  
\put(8,16){\includegraphics[width=8cm]{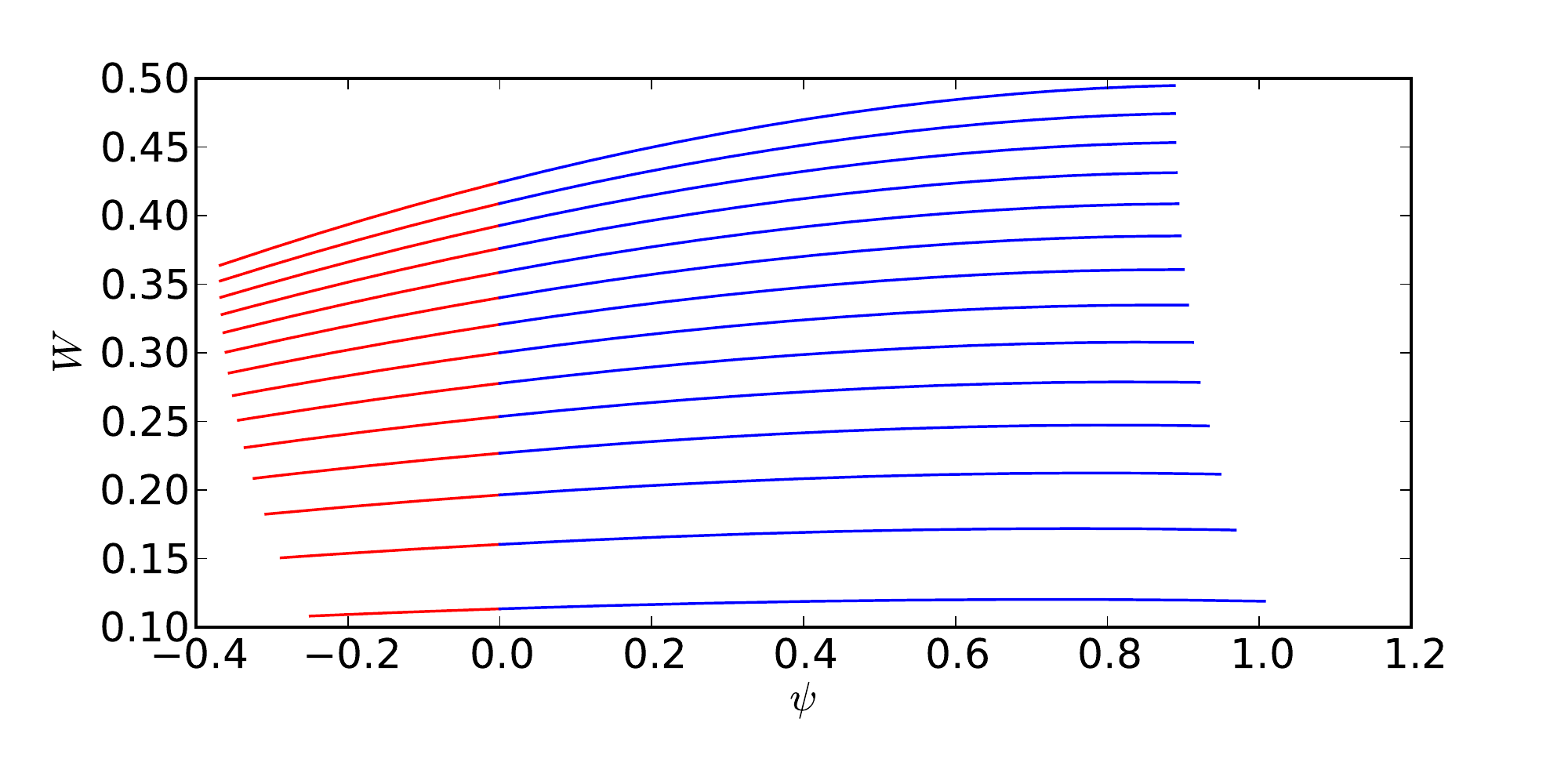}}  
\put(8,12){\includegraphics[width=8cm]{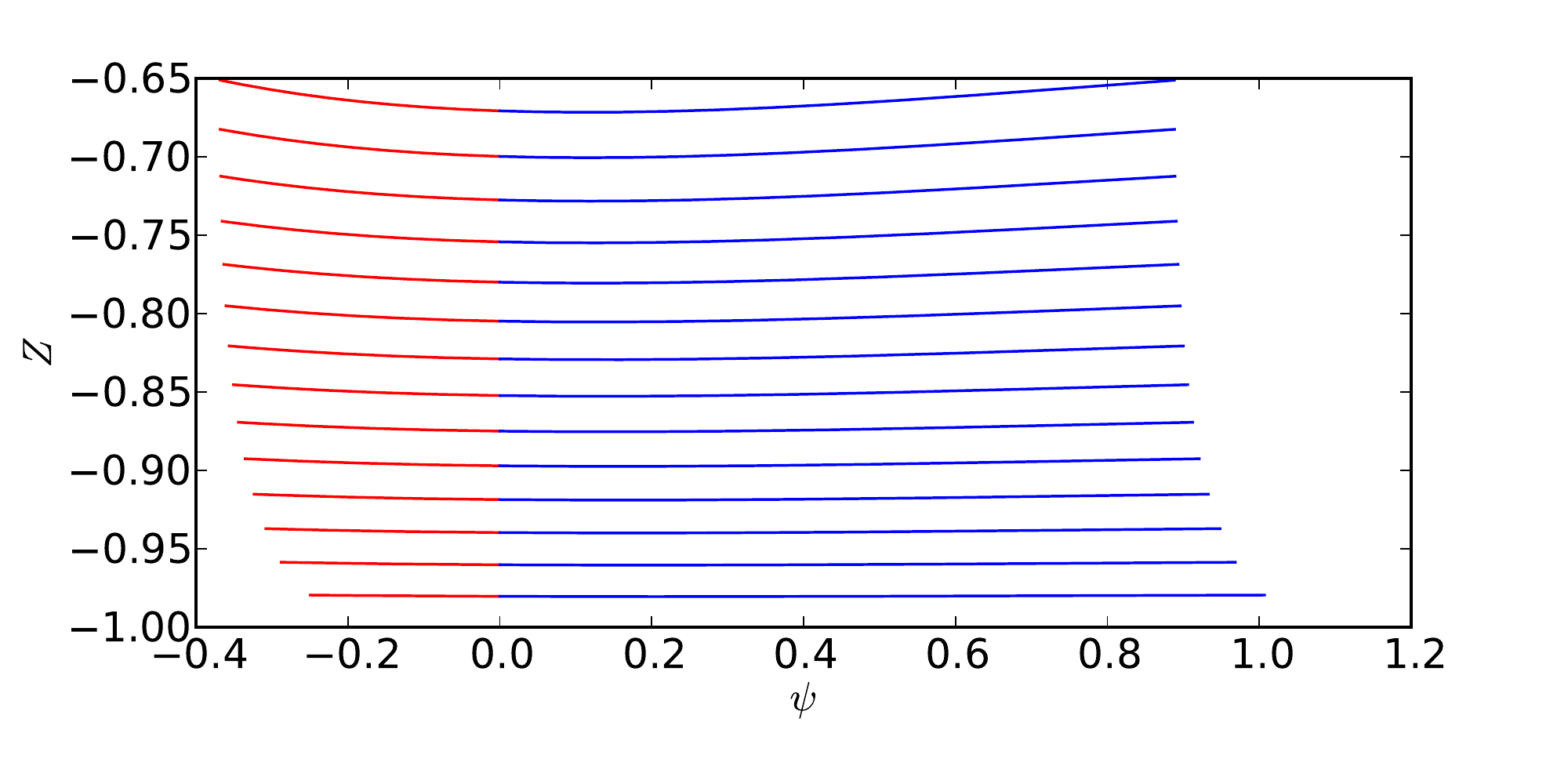}}  
\put(8,8){\includegraphics[width=8cm]{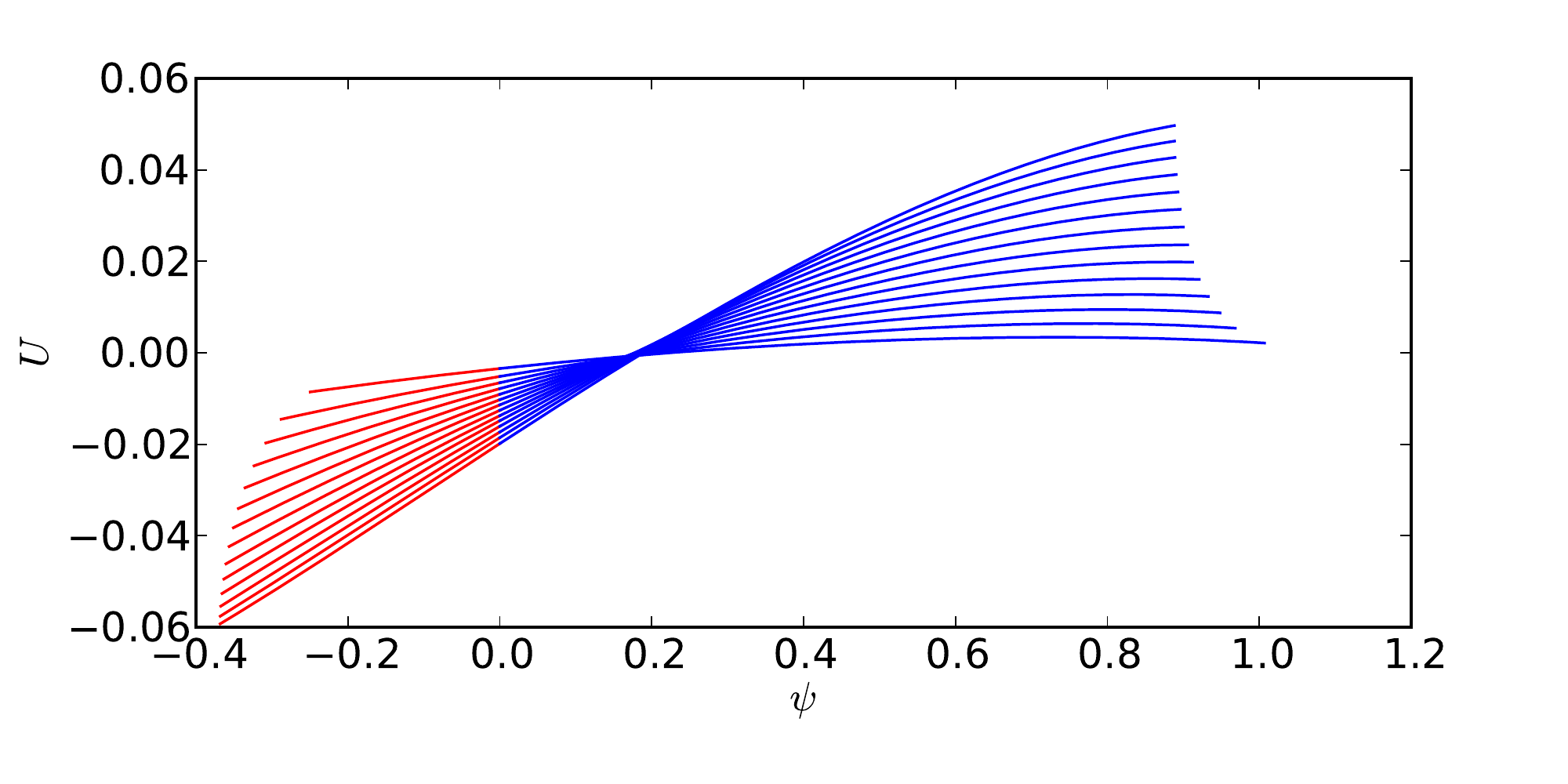}}  
\put(8,4){\includegraphics[width=8cm]{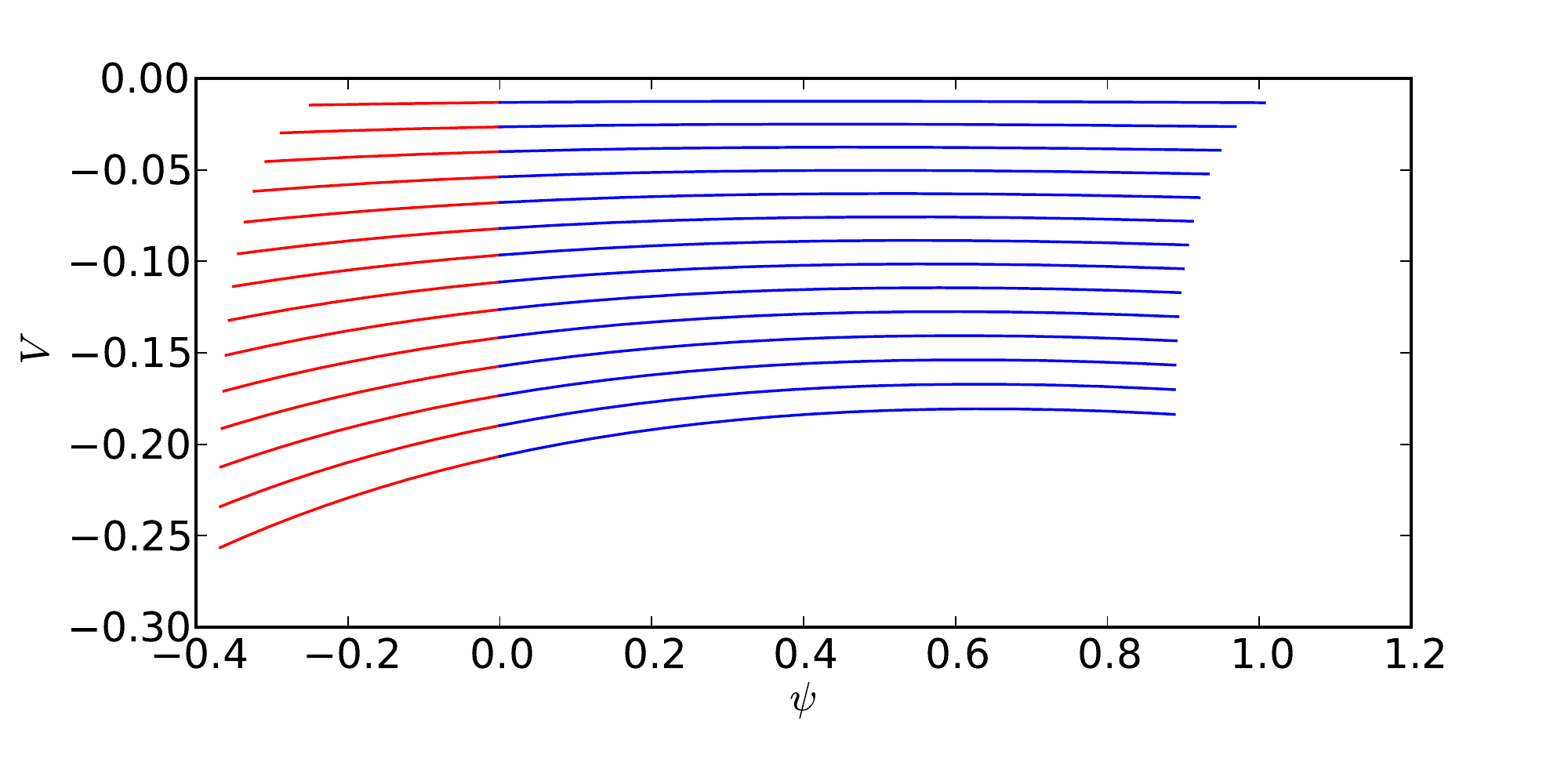}}  
\put(8,0){\includegraphics[width=8cm]{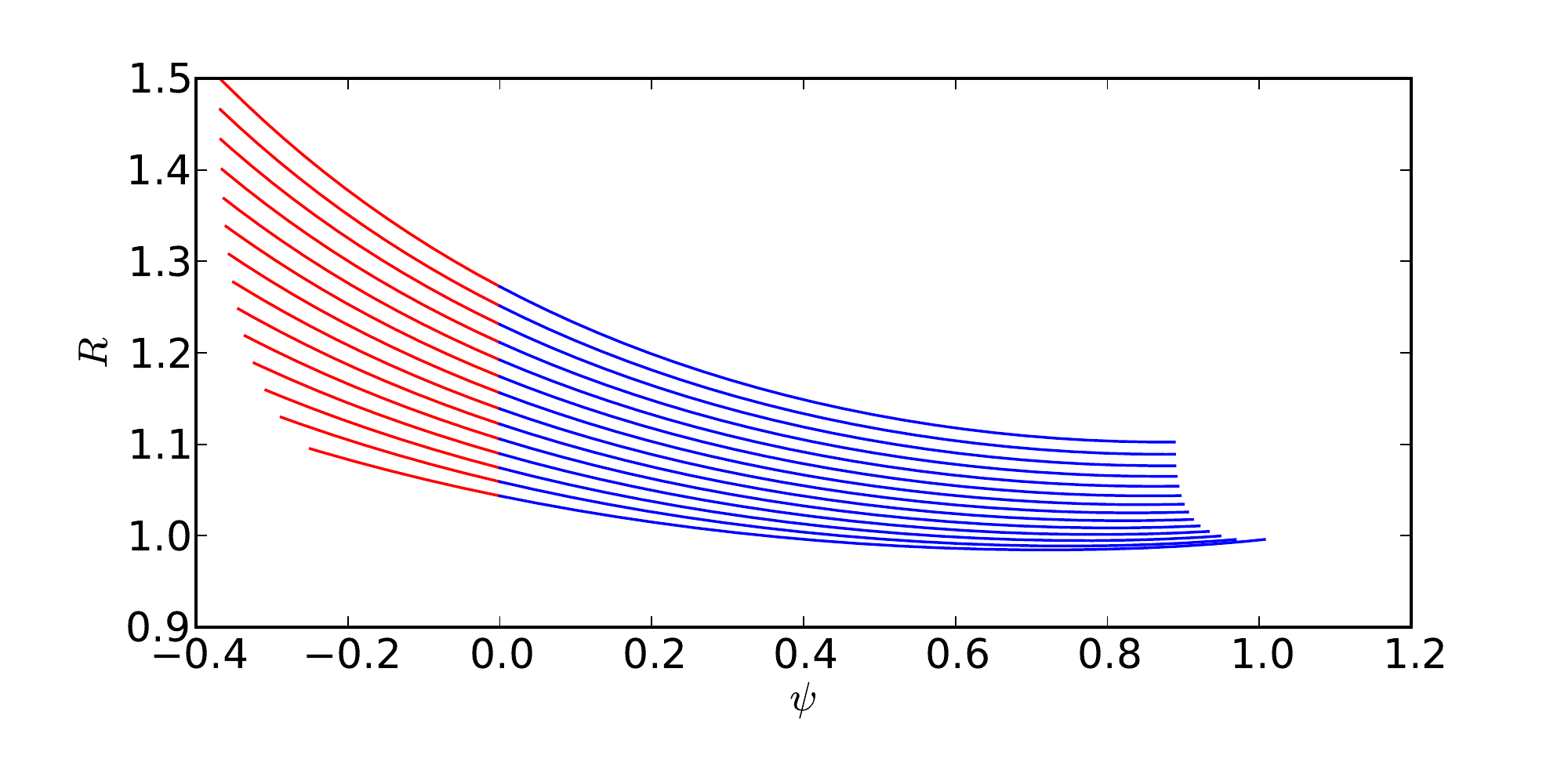}}  
\end{picture}
\caption{Same as Fig.~\ref{mode_m2} for the modes $m=5$.}
\label{mode_m5}
\end{figure*}

For compleness Fig.~\ref{mode_m5} shows the $m=5$ solutions for the same
temperatures as in Fig.~\ref{mode_m2}. Altogether, the solutions have
similar shape as  the $m=2$ mode but present less variations,
therefore as will be seen in Sect.~\ref{stress_sec}, they lead to
smaller mass fluxes than the $m=2$ mode.

\section{\label{app:approx}Analytical expansion in the low temperature limit}

Here we present an analytic expansion of Eqs.~(\ref{ef_new_od1}-\ref{ef_new_od2}) 
that is valid in the low temperature limit. Since we solved these equations 
numerically, the aim is more to make the various dependence more explicit rather than 
obtaining accurate expressions. The approximated expressions are typically good
at low temperature and at large $m$. 

Let us expand the perpendicular and parallel velocity, $W$ and $Z$ as
\begin{eqnarray}
\nonumber
W(\psi) &=& \sqrt{T_0} ( 1 +  w(\psi) ), \\
Z(\psi) &=& Z_c +  z(\psi),
\label{expand}
\end{eqnarray}
where we will assume that $w<<1$ and $z << Z_c$. 
Since at the critical point $w=0$ and $z=0$, this assumption 
is clearly verified in the vicinity of  the critical point and therefore 
particularly in the limit of large $m$.
Moreover as discussed previously the spiral parameter, $B=\tan(\theta)$,
is on the order of $1/\sqrt{T_0}$. Thus we write $B=b/\sqrt{T_0}$.
Finally, we restrict the calculation to the case $n=1/2$.

Plunging these expressions into  Eqs.~(\ref{ef_new_od1}-\ref{ef_new_od2}), we get

\begin{eqnarray}
\label{expand}
w w ' &=& A_1 w + A_2 z, \\
\nonumber
z ' &=& B_0 + B_1 w + B_2 z, \\
\nonumber
A_1 &=& {1 \over 4 (T_0 + b^2)}  \left(  (2 b -Z_c) \sqrt{T_0} \right) \simeq {3 \sqrt{T_0} \over 4}, \\
\nonumber
A_2 &=& {1 \over 4 (T_0 + b^2)}  \left( {4 Z_c \over \sqrt{T_0}} - \sqrt{T_0} \right) \simeq {-1 \over  \sqrt{T_0}}, \\
\nonumber
B_0  &=& { \sqrt{T_0} \over (T_0 + b^2)} \left(  {Z_c^2 \over 2} - {b Z_c \over 2} - 1 + {5 \over 2} T_0 \right)
\simeq   { (b-1) \sqrt{T_0}   } , \\
\nonumber
B_1  &=& { \sqrt{T_0} \over (T_0 + b^2)} \left( 2 \sqrt{T_0} - {b \over 2} Z_c  \right) \simeq { \sqrt{T_0} \over 2} , \\
\nonumber
B_2  &=& { \sqrt{T_0} \over (T_0 + b^2)} \left(Z_c - {b \over 2}  \right) \simeq { - 3 \sqrt{T_0} \over 2}, 
\end{eqnarray}
where to get the simplified expressions, we have used that in the limit of low $T_0$
\begin{eqnarray}
Z_c = - 1 + T_0 {1 + 4 b \over 4 b} \simeq -1,
\label{crit_expand}
\end{eqnarray}
while 
\begin{eqnarray}
b  \simeq   1.
\label{b_expand}
\end{eqnarray}

To get an approximation of Eqs.~(\ref{expand}) we expand $w$ and $z$ to the second order in $\psi - \psi_c$,
\begin{eqnarray}
\nonumber
w (\psi) &=& a_1 (\psi-\psi_c) + a_2 (\psi-\psi_c)^2, \\
\nonumber
z (\psi) &=& b_1 (\psi-\psi_c) + b_2 (\psi-\psi_c)^2, 
\label{wz_expand}
\end{eqnarray}
The coefficents $a_{1,2}$ and $b_{1,2}$ are solutions of the following 
equations 
\begin{eqnarray}
\nonumber
a_1^2 &=& A_1 a_1 + A_2 b_1, \\
\nonumber
3 a_1 a_2 &=& A_1 a_2 + A_2 b_2, \\
\nonumber
b_1 &=& B_0, \\
2 b_2 &=& B_1 a_1 + B_2 b_1.
\label{wz_equ}
\end{eqnarray}
which leads to 
\begin{eqnarray}
a_1 &=& {1 \over 2} \left( A_1 + \sqrt{A_1^2 + 4 A_2 B_0} \right)    \simeq \sqrt{1-b}, \\
\nonumber
a_2 &=& {b_2 A_2 \over 3 a_1 - A_1}  \simeq  -{1 \over 12} {1 \over 1 - {1 \over 4} \sqrt{T_0 \over 1 -b }}, \\
\nonumber
b_1 &=&  B_0  \simeq - \sqrt{T_0} (1-b), \\
b_2 &=&   {1 \over 2} (a_1 B_1 + B_2 B_0) \simeq  {1 \over 4} \sqrt{T_0} \sqrt{1-b} , 
\label{wz_expand_sol}
\end{eqnarray}

At this stage we have two free parameters, $\psi_c$ and $b$. They are determined by the 
two boundary conditions as given by Eqs.~(\ref{RK2}-\ref{RK_W}), which in the limit
$w <<1$ leads to
\begin{eqnarray}
\nonumber
w\left( {\pi \over m} - \psi_c \right) &=& - w\left( -{\pi \over m} - \psi_c \right), \\
z\left( {\pi \over m} - \psi_c \right) &=& z\left( -{\pi \over m} - \psi_c \right).
\label{boundary_dev}
\end{eqnarray}
Combining them with Eqs.~(\ref{wz_expand_sol}), we obtain
\begin{eqnarray}
\nonumber
\psi_c =  {b_1 \over 2 b_2} \simeq   - 2 \sqrt{1-b}, \\
a_1 \psi _c = a_2 \left( \left( {\pi \over m} \right)^2 + \psi_c^2 \right).
\label{res_bound}
\end{eqnarray}
Combining these two last equations, we get a non-linear equation of $b$, 
which can be easily solved using a standard root finder. 
Once we get $b$, all quantities are known. 
A comparison between the numerical solutions and the approximated ones 
is displayed in Fig.~\ref{comparison}.

\setlength{\unitlength}{1cm}
\begin{figure} 
\begin{picture} (0,12)
\put(0,6){\includegraphics[width=8cm]{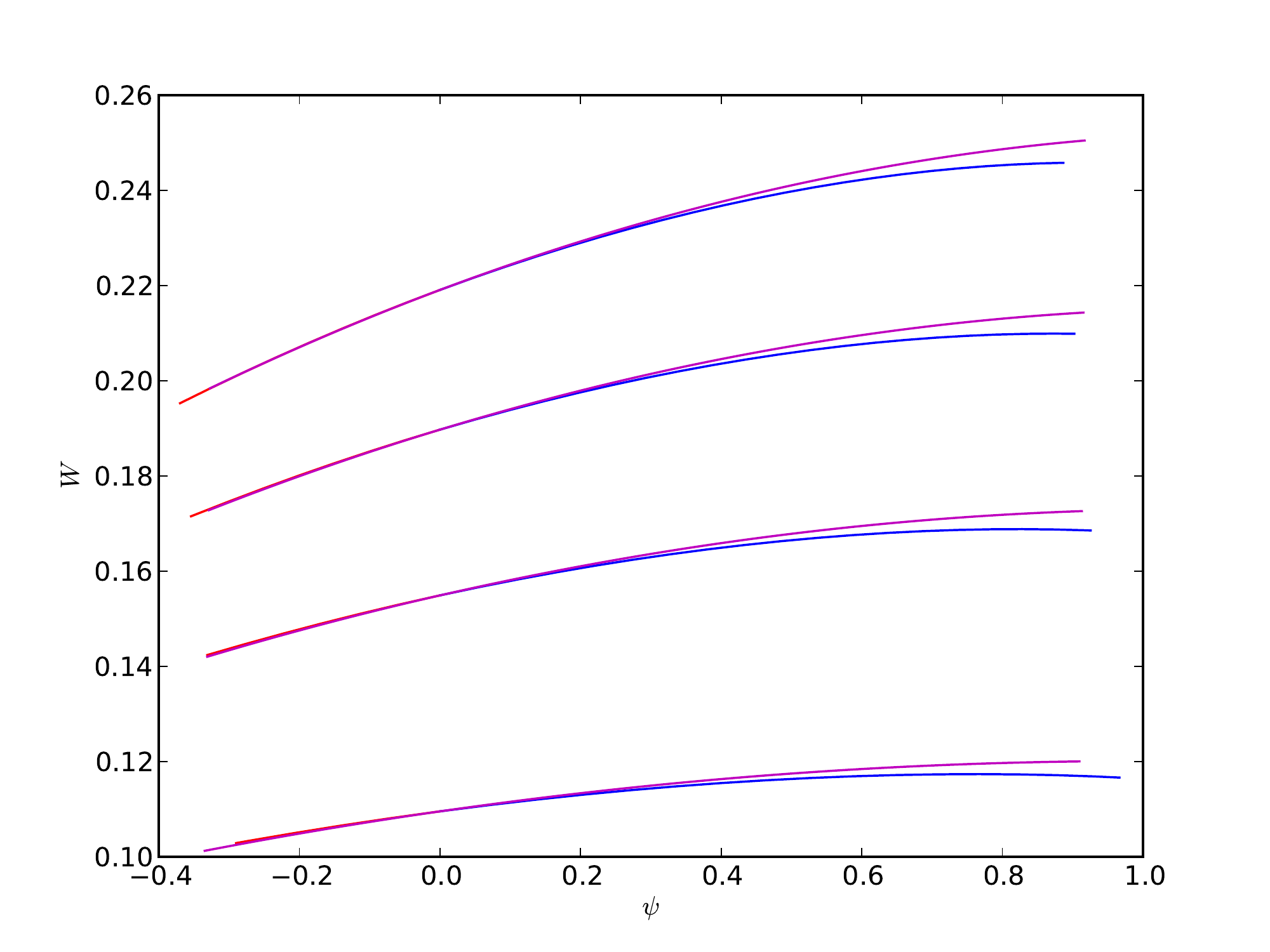}}  
\put(0,0){\includegraphics[width=8cm]{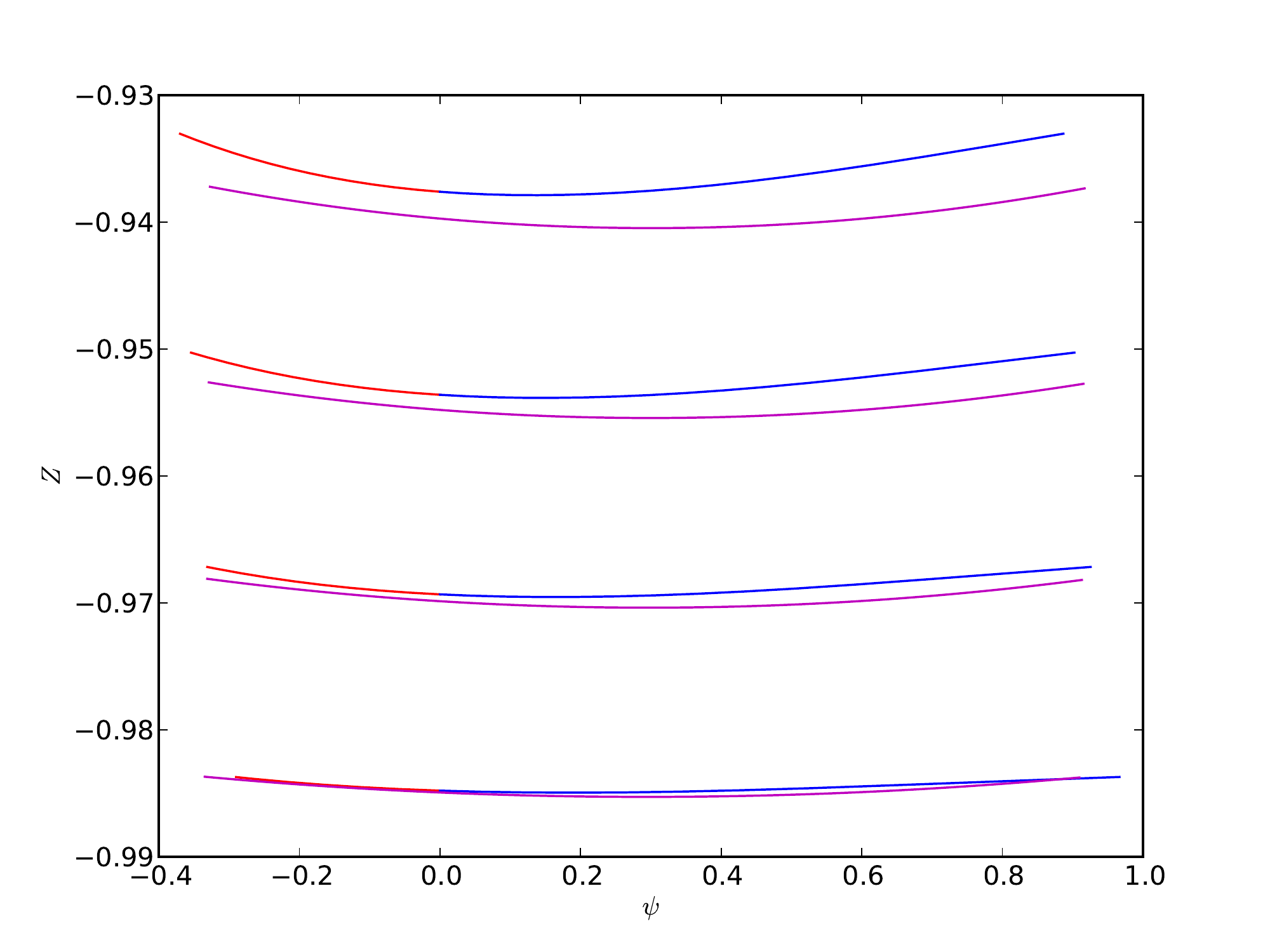}}  
\end{picture}
\caption{Comparison between the numerical solutions (red and blue curves) obtained
numerically and the approximated ones presented in this appendix (magenta curves)
for $m=5$ and $T_0$=0.048, 0.036, 0.024 and 0.012. As can be seen they are 
quite close demonstrating the validity of the approximation. }
\label{comparison}
\end{figure}


\bibliographystyle{aa}
\bibliography{biblio}

\end{document}